\documentclass[final,5p,times,twocolumn,authoryear]{elsarticle}
\usepackage{tabularx} 
\usepackage{amsmath}  
\DeclareMathOperator{\arccot}{arccot}
\usepackage{graphicx} 
\usepackage{physics}
\usepackage[title]{appendix}
\usepackage[final]{hyperref} 
\usepackage{bm}
\usepackage{bbm}
\usepackage{amssymb}
\usepackage{color}
\usepackage{natbib}
\setcitestyle{numbers,square}
\hypersetup{
	colorlinks=true,       
	linkcolor=blue,        
	citecolor=blue,        
	filecolor=magenta,     
	urlcolor=blue         
}

\journal{Science Bulletin}

\begin{document}
\begin{frontmatter}

\title{Spectral parameters of the $\rho$ resonance from lattice QCD
\\[1.5em]
\includegraphics[scale=0.3]{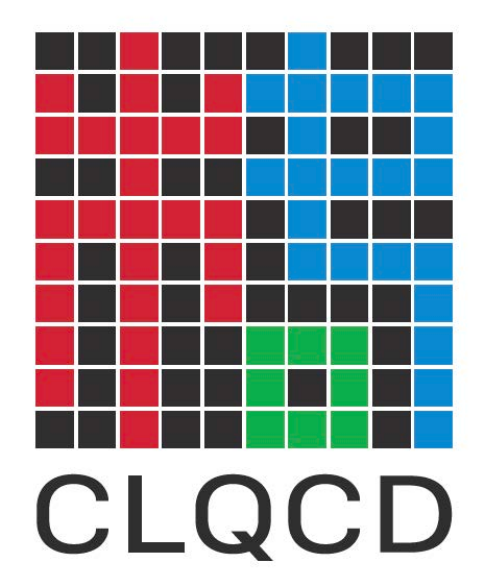}}
\author[1]{Zhengli Wang}
\ead{wangzhengli@ucas.ac.cn}

\author[7]{Derek B. Leinweber}
\author[2,3,4]{Chuan Liu}
\author[5,1]{Liuming Liu\corref{cor1}}
\ead{liuming@impcas.ac.cn}

\author[5,1]{Peng Sun}
\author[7]{Anthony W. Thomas}
\author[1,8]{Jia-jun Wu\corref{cor1}}
\ead{wujiajun@ucas.ac.cn}
\author[5,1]{Hanyang Xing}
\author[1]{Kang Yu}

\cortext[cor1]{Corresponding author}
\address[1]{\textit{ University of Chinese Academy of Sciences, School of Physical Sciences, Beijing 100049, China}}
\address[2]{\textit{ School of Physics, Peking University, Beijing 100871, China}}
\address[3]{\textit{ Center for High Energy Physics, Peking University, Beijing 100871, China}}
\address[4]{\textit{ Collaborative Innovation Center of Quantum Matter, Beijing 100871, China}}
\address[5]{\textit{ Institute of Modern Physics, Chinese Academy of Sciences, Lanzhou 730000, China}}
\address[7]{\textit{ Special Research Centre for the Subatomic Structure of Matter (CSSM), \\ 
Department of Physics, University of Adelaide, Adelaide, South Australia 5005, Australia}}
\address[8]{%
Southern Center for Nuclear-Science Theory (SCNT), Institute of Modern Physics,
Chinese Academy of Sciences, Huizhou 516000, Guangdong Province, China
}%

\begin{abstract}
We present a lattice QCD investigation of the $\rho$ resonance using nine $N_f = 2 + 1$ Wilson-Clover ensembles with three lattice spacings and various pion masses ranging from $135$ to $320$ MeV. 
For each ensemble, a large number of finite volume energy levels are determined and the energy dependence of the phase shift obtained from L\"uscher's finite volume method. The mass and width of the $\rho$ resonance are then extracted by assuming the Breit-Wigner form. 
The mass and width are extrapolated to the physical pion mass and continuum limit ($\mathcal{O}(a^2)$) using a linear function of $a^2$ and $m^2_\pi$. 
The extrapolated values for the mass and width in the Breit-Wigner form are $(m_\rho,\,\Gamma_\rho) = (781.6\pm10.0,\, 146.5\pm 9.9)$ MeV, which are in good agreement with experiment.
An alternative method of analysis, based on Hamiltonian effective field theory, involves directly fitting the lattice energy levels and accounting for the quark mass dependence of the hadronic loop diagrams which yield the leading and  next-to-leading non-analytic behaviour. This approach also yields consistent $\rho$ parameters at the physical point.
This represents the most precise determination to date of the mass and width of a hadron which is unstable under strong decay, achieved through comprehensive lattice QCD calculations and methods of analysis.
\end{abstract}


\end{frontmatter}

\section{Introduction}

The fundamental theory of the strong interaction is Quantum Chromodynamics (QCD).
In the high-energy regime, the property of asymptotic freedom allows for precise tests of QCD through perturbative calculations. 
However, the confinement property in the low-energy regime presents significant challenges in understanding the non-perturbative aspects of strong interactions.
These non-perturbative properties manifest in the mass spectra, the decay properties of various hadronic states and the scattering amplitudes of hadronic reactions.
Because of the short lifetimes of many particles which are stable under strong interactions but decay through weak interactions, such as $\Lambda$ baryons, $D$ meson, $J/\psi$ meson, etc., it is difficult to produce beams of these particles in experiments.
As a result, many two-body scattering processes cannot be realized in experiments, and the available data on hadron scattering is extremely limited.

On the other hand, after more than half a century of development, lattice QCD (LQCD) has become an important method for studying strong interactions. 
Through numerical methods, LQCD provides an \textit{ab initio} approach to calculate hadron scattering amplitudes.
LQCD has achieved great success in the study of stable particles. 
Resonant particles with widths ranging from tens to hundreds of MeV, such as the $\rho$, $\Delta$, $f_0(980)$, etc., have also been investigated extensively in LQCD. 
Among them, the $\rho$ meson is the lightest vector meson, representing the ground state in the quark model and predominantly decaying into the $\pi\pi$ final state. 
Given its significance as a prototypical resonant particle in strong interactions, the $\rho$ meson serves as an ideal object for a precise determination of hadronic properties through LQCD.

Many LQCD groups have studied the finite-volume energy spectrum of the $\rho$ meson under various conditions and have determined the pole position of the $\rho$ meson~\cite{Oller:1998hw,McNeile:2002fh,Bruns:2004tj,Allton:2005fb, Armour:2005mk, Michael:2006hf, CP-PACS:2007wro, Gockeler:2008kc, Hanhart:2008mx, Nebreda:2010wv, Pelaez:2010fj, PACS-CS:2010dxu, Budapest-Marseille-Wuppertal:2010gis, Baron:2010bv, EuropeanTwistedMass:2010voq, Feng:2010es, Lang:2011mn, CS:2011vqf, Guo:2011pa, Guo:2012yt, Chen:2012rp, Pelissier:2012pi, Dudek:2012xn, Bruns:2013tja, Albaladejo:2013bra, Metivet:2014bga, Bali:2015gji, Bolton:2015psa, Wilson:2015dqa, Sun:2015enu, Guo:2015dde, Fu:2016itp, Guo:2016zos, Bulava:2016mks, Hu:2016shf, Alexandrou:2017mpi, Andersen:2018mau, Giusti:2018mdh, ExtendedTwistedMass:2019omo, Boyle:2024hvv, Wu:2014vma, Yan:2024gwp}.
In these studies, a comprehensive research methodology has been established.
It is worth mentioning that for these unstable hadronic states, the resonance energy is distributed among all finite volume eigenstates corresponding to the quantum numbers of the hadron~\cite{Wu:2016ixr}. 
Therefore, in order to accurately extract resonance properties from lattice energy levels, it is necessary to identify all energy levels near the mass of the hadronic state. 

For the $\rho$ meson, two types of operators must be established, the $q\bar{q}$($q=u, d$) operator and the two-hadron operator of $\pi\pi$, where different momentum combinations of the $\pi\pi$ correspond to different operators. 
Computing the correlation function matrix of such large operator bases is made efficient using the distillation quark smearing method~\cite{HadronSpectrum:2009krc}, which also renders simple the inclusion of quark annihilation diagrams with a high degree of statistical precision.
After obtaining the correlation function matrix, different energy levels can be obtained through the Generalized Eigenvalue Problem (GEVP) method.

In order to increase the number of finite-volume energy levels, it is possible to compute the spectrum not only in the rest frame of the $\rho$ meson but also in moving frames. 
Since in the energy region of the $\rho$ meson, there is only one open channel, $\pi\pi$, the L\"uscher formula~\cite{Luscher:1985dn, Luscher:1986pf, Luscher:1990ux} and its generalization to the moving system~\cite{Rummukainen:1995vs, Kim:2005gf, Feng:2010es, Gockeler:2012yj, Li:2024zld} can be directly used to convert each energy level into the phase shift of $\pi\pi$ scattering at the corresponding energy.
Finally, the pole position of the $\rho$ meson can be obtained by fitting these phase shifts.

Although a significant amount of lattice data on the $\rho$ meson has been accumulated, the data set lacks uniformity. For example, different fermion actions have been used by these lattice groups, rendering different discretization effects in these lattice data. 
Additionally, the strange quark mass is also different among these groups. 
In fact, it is not possible to analyze the pion mass dependence of the $\rho$ meson within the framework of chiral effective field theory because of these differences, as detailed in Ref.~\cite{Yu:2023xxf}. 
Furthermore, many previous lattice calculations were performed at a specific lattice spacing, so the effects of the finite lattice spacing were only roughly estimated.
Although some earlier studies did use several different lattice spacings to study the $\rho$, they did not have physical pion mass data and did not extract the influence of lattice spacing at the $\cal{O}$ $(a^2)$ level. 
Therefore, there is an urgent need to create a consistent set of configurations with the same fermion action but different lattice spacings and pion masses, particularly including the physical pion mass, in order to systematically study the properties of the $\rho$ meson.

In this work, we utilize 9 sets of $2 + 1$ flavor Wilson-Clover gauge configurations, with three lattice spacings and various pion masses ranging from $135$ to $320$ MeV, to simultaneously extrapolate the $\rho$ resonance parameters to the physical pion mass and continuum limit. 
For each ensemble, around 30 energy levels are obtained using operators in both the rest frame and moving frames with total momentum $\vec{P} = (0,0,1)$, $(0,1,1)$, $(1,1,1)$ and $(0,0,2)$ with lattice unit.

Subsequently, the $p$-wave $\pi\pi$ scattering phase shifts at these energies are obtained using the L\"uscher formula.
We find that the phase shifts exhibit a fast transition from $0^\circ$ to $180^\circ$, indicating the presence of a resonance. 
The energy where the transition happens is approximately the $\rho$ meson mass.

By fitting the phase shift data near the $\rho$ meson mass using the most commonly used Breit-Wigner (BW) form, six groups of masses and widths for the $\rho$ meson are determined at three lattice spacings and various pion masses. 
Finally, a linear extrapolation of the pion mass and lattice spacing dependence is performed to obtain the $\rho$ meson mass and width at the physical pion mass in the continuum limit. We find  $(m_\rho,\,\Gamma_\rho) = (\,781.6\,(10.0),\, 146.5\,(9.9)\,)$ MeV, as well as the pole position $Z_{\text{pole}} = 768.1\,(10.0)-i\,70.5\,(4.9)$ MeV, which are in excellent agreement with the experimental results~\cite{ParticleDataGroup:2024cfk}. 

The analysis is repeated a second time using an alternative finite volume
approach, namely Hamiltonian effective field theory (HEFT).
While HEFT incorporates the L\"uscher formalism, it is also able to describe the quark mass dependence of the hadronic energy levels and their associated scattering observables. 
As such, this approach provides a robust formalism for accurately describing the non-analytic quark mass dependence of the $\rho$-meson spectral properties. 
Using our most comprehensive analysis, incorporating the $\pi\omega$ channel, we
find $(m_\rho,\,\Gamma_\rho) = (\,782.0\,(13.5),\, 155.0\,(12.0)\,)$ MeV. This and the corresponding pole position, $m^{\text{pole, ext(B)}}_\rho =777.0\,(15.0)-i60.0\,(6.0)$ MeV, agree well with experiment and the more traditional L\"uscher-BW formalism, as shown in the previous paragraph.

This paper is organized as follows. 
In Sec.~\ref{sec:tools}, we review the details of our lattice setup, operator construction,  variational analysis, and the formulas involved in the phase shift, pole position and extrapolation analyses.
In Sec.~\ref{sec:result} we show the extracted finite-volume spectra, phase shifts, and pole positions for each pion mass and lattice spacing.  
Finally we obtain the pole position of the $\rho$ meson at the physical pion mass in the continuum limit via linear extrapolation.
Here we use two methods to make the extrapolation.
One uses the L\"uscher formula and the assumption of a BW form to extrapolate mass and width The other approach is based on Hamiltonian effective field theory (HEFT) which simultaneously describes the finite volume and quark mass dependence of the results.  
Both methods provide consistent values for the pole position of the $\rho$ meson.
%
At last we give a summary in Sec.~\ref{sec:Summary}.

\section{Computational methodology}
\label{sec:tools}

\subsection{Lattice setup}
\label{sec:Lattice_action}
The results presented in this paper are based on the gauge configurations generated by the CLQCD collaboration with 2+1 dynamical quark flavors, employing the tadpole-improved Symanzik-improved gauge action and Clover fermion action~\cite{CLQCD:2023sdb}. 
We use nine ensembles with three lattice spacing values: $a$ = 0.105~fm, 0.077~fm and 0.052~fm, and various pion masses ranging from 135~MeV to 320~MeV. 
The detailed information regarding these ensembles is provided in Table~\ref{tab:lattice}. 
Among these ensembles, C24P29/C32P29, C32P23/C48P23 and F32P30/F48P30 are three pairs that share the same lattice spacing and pion mass but have different volumes.
The ensemble C48P14 has the physical pion mass. 
The diverse parameters across these ensembles facilitate a robust extrapolation of our results to the physical pion mass and continuum limit. 
The $K$ meson mass $m_K$ for all ensembles are similar and close to the physical value, therefore we neglect the influence of the strange quark mass in the subsequent analysis.  

The quark propagators are computed using the distillation quark smearing method~\cite{HadronSpectrum:2009krc}, which improves precision, enables efficient calculation of the correlation function matrix of many interpolating operators, and allows for the computation of quark annihilation diagrams. 
The smearing operator is composed of a small number ($N_{ev}$) of the eigenvectors associated with the $N_{ev}$ lowest eigenvalues of the three-dimensional Laplacian defined in terms of the HYP-smeared gauge field. 
The number of eigenvectors $N_{ev}$ is 100 for the ensembles with $L$ = 24 and 32, and 200 for the ensembles with $L$=48. 
The source time is put on every time slice to improve statistics. 

\begin{table*}[htbp]
	\centering
	\caption{\label{tab:lattice} Parameters of the ensembles and the propagators. The listed parameters are the lattice spacing $a$, lattice size $(L/a)^3\times T/a$, dimensionless quark mass parameters for light and strange quarks $\tilde{m}^{\rm b}_{l,s}$, renormalized quark masses $m^R_{l,s}$, the mass of pion/kaon $m_{\pi}$/$m_{K}$, the value of $m_\pi L$, the number of configurations $N_{\mathrm{cfgs}}$ and the number of eigenvectors $N_{ev}$ used in the distillation method.}
	\resizebox{2\columnwidth}{!}{
		\begin{tabular}{l | rrrrr | rrr | r}
			& C24P29 &  C32P29 & C32P23 & C48P23 & C48P14 & F32P30  & F48P30  & F48P21 & H48P32 \\   
			\hline    
			$a$ (fm) & \multicolumn{5}{c|}{0.10530(18)} & \multicolumn{3}{c|}{0.07746(18)}  & \multicolumn{1}{c}{0.05187(26)} \\
			$\tilde{L}^3\times \tilde{T}$ & $24^3\times 72$ & $32^3\times 64$ & $32^3\times 64$ & $48^3\times 96$ & $48^3\times 96$ & $32^3\times 96$ & $48^3\times 96$ & $48^3\times 96$ & $48^3\times 144$ \\
			\hline 		
			$\tilde{m}^{b}_l$ & -0.2770 & -0.2770 & -0.2790 & -0.2790 & -0.2825 & -0.2295 &  -0.2295 & -0.2320 & -0.1850\\	  
			$\tilde{m}^{b}_s$  & -0.2400 & -0.2400 & -0.2400 & -0.2400 & -0.2310 & -0.2050 & -0.2050 & -0.2050 & -0.1700\\	
			\hline 	
			$m^R_{l}$ (MeV) & 16.94(12)       &  17.35(11)   &	 10.55(11)  &	10.27(10)	&   3.638(83)   &	18.54(12)	 & 18.511(92) & 8.59(08)   &	19.42(05)\\
			$m^R_{s}$ (MeV) & 87.46(10)
			& 88.16(10) & 84.48(07) & 84.79(04)
			& 103.15(05) & 93.23(11) & 93.05(08) 
			& 90.43(08)& 95.61(04)\\       
			\hline 	
			$m_{\pi}$\ \ (MeV) & 292.7(1.2)	  &  292.4(1.1)	 &   228.0(1.2)	&    225.6(0.9) &	135.5(1.6)  &   303.2(1.3)	 & 303.4(0.9) & 207.2(1.1)	 &  317.2(0.9)\\
			$m_{K}$ (MeV) &509.4(1.1)   &509.0(1.1)    & 484.1(1.0)   &484.1(1.3)    & 510.0(1.0)  & 524.6(1.8)   &523.6(1.4)   &  493.0(1.4)  &536.1(3.0) \\
			\hline
			$m_\pi L$ & 3.75 & 5.01 & 3.91 & 5.79 & 3.56 & 3.81 & 5.72 & 3.91 & 4.06 \\
			\hline
			$N_{\mathrm{cfgs}}$ & 880 & 430 & 450 & 265 & 260 & 370 & 200 & 220 & 270 \\
			\hline
			$N_{ev}$ &100 & 100 &100 &200 &200 &100 &200 &200 &200 
		\end{tabular}
	}
\end{table*}

\subsection{Interpolating operators}
\label{sec:Interpolators}
The $SO(3)$ rotational symmetry of the continuum is broken to the octahedral group,  $O_h$, on a finite lattice and is further reduced to the little groups when going to the moving frames. 
Consequently, the interpolating operators should transform according to the irreducible representations (irreps) of the $O_h$ group and the little groups. 
We will consider the operators in the rest frame as well as the moving frames with momenta $\vec{P} = (0,0,1)$, $(0,1,1)$, $(1,1,1)$ and $(0,0,2)$. 
For the $\rho$ meson with $J^P = 1^{-}$, the relevant irreps are listed in Table~\ref{tab:irreps}. 

\begin{table}[htbp]
\begin{centering}
\caption{The little groups $LG(\vec{P})$ for all frames used in this work and the decomposition of $J=1$ into the irreps.}\label{tab:irreps}
\hspace{1.5cm}
\begin{tabular}{ccc}
\hline
\hline
$\vec{P}$  & $LG(\vec{P})$ &irreps \\
\hline
(0,0,0) &$O_h$  &$T_1^-$ \\
\hline
(0,0,1) &$C_{4v}$ & $A_1 \oplus E$  \\
\hline
(0,1,1) &$C_{2v}$ & $A_1 \oplus B_1 \oplus B_2$ \\
\hline
(1,1,1) &$C_{3v}$ &$A_1 \oplus E$ \\
\hline
(0,0,2) & $C_{4v}$ & $A_1 \oplus E$ \\
\hline
\hline
\end{tabular}
\end{centering}
\end{table}

In Ref.~\cite{Dudek:2012xn}, which considered the $\rho$ resonance in $\pi\pi$ elastic scattering, it was found that in order to reliably extract the complete low-energy spectrum of finite volume eigenstates, operators resembling both single hadrons and multi-hadrons must be included in the basis.
For the single meson operator, we use the local quark bilinear operator: 
\begin{equation}
\mathcal{O}_{\rho, i}\,(\vec{x}, t\,) =\frac{1}{\sqrt{2}}(\,\bar{u} \gamma_i u\,(\vec{x}, t\,) - \bar{d} \gamma_i d\,(\vec{x}, t\,)\,), 
\end{equation}
where $u$ and $d$ represent up and down quark fields and $i=\{1,2,3\}$ is used for the Lorentz indices. 
The momentum projected operator is defined as $\mathcal{O}(\vec{P}, t) = \sum_{\vec{x}} e^{-i\vec{p}\cdot\vec{x}}\mathcal{O}(\vec{x}, t)$. 
The operators in the irreps of lattice groups are appropriate linear combinations of $\mathcal{O}_{\rho, i}$. 
We follow the procedure described in Refs.~\cite{Dudek:2010wm,Thomas:2011rh}.

In the rest frame, the single meson operators with definite spin $J$ and $z-$component $M$ are constructed as follows:
\begin{align}
\mathcal{O}^{J=1,M=1} &= -\frac{1}{\sqrt{2}}(\mathcal{O}_{\rho, 1} + i\, \mathcal{O}_{\rho, 2}),  \nonumber \\
\mathcal{O}^{J=1,M=0} &= \mathcal{O}_{\rho, 3}, \\
\mathcal{O}^{J=1,M=-1} &= \frac{1}{\sqrt{2}}(\mathcal{O}_{\rho, 1} - i\, \mathcal{O}_{\rho, 2}), \nonumber
\end{align} 
and then subduced to the $T_1^-$ irrep of the $O_h$ group. 
The subduction is trivial.

For the moving frames, we first construct the helicity operator 
$\mathcal{O}^{J, \lambda} = \sum_M \mathcal{D}^{(J)*}_{M\lambda}(R) \, \mathcal{O}^{J,M}(\vec{p})$,
%
where $\lambda$ is the helicity, $\mathcal{D}^{(J)}_{M\lambda}(R)$ is the Wigner-$\mathcal{D}$ matrix and $R$ rotates $(0,0,|\vec{p}|)$ to $\vec{p}$. 
The operators in the little groups are obtained by subducing the helicity operators, with the subduction coefficients provided in Ref.~\cite{Thomas:2011rh}.

The $\pi\pi$ two-meson operators can be generally written as:
\begin{equation} \label{eq:two-meson}
\left(\pi\pi\right)_{\vec{P},\,\Lambda,\,\mu}^{[\vec{k}_1,\vec{k}_2]} = \sum_{\substack{\vec{k}_1 \in \{\vec{k}_1\}^\star  \\ \vec{k}_2 \in \{\vec{k}_2\}^\star \\ \vec{k}_1 + \vec{k}_2 =\vec{P} }} \mathcal{C} \left(\vec{P},\Lambda,\mu;\vec{k}_1 ; \vec{k}_2\right) \times\left( \pi^+(\vec{k}_1) \pi^-(\vec{k}_2) - \pi^-(\vec{k}_1) \pi^+(\vec{k}_2)\right),
\end{equation}
where $\pi^+ = \bar{d} \gamma_5 u$ and $\pi^- = \bar{u} \gamma_5 d$ are single pion operators, $\mathcal{C}$ is a Clebsch-Gordan coefficient for $\Lambda_1 \otimes \Lambda_2 \to \Lambda$ with $\Lambda_{1,2} = A_1^-$ of $O_h^D$ if $\vec{k}_{1,2}=\vec{0}$ and $\Lambda_{1,2} = A_2$ of the little group $LG(\vec{k}_{1,2})$ if $\vec{k}_{1,2}\neq\vec{0}$, $\Lambda$ is an irrep of  $LG(\vec{P})$ and $\mu$ is the row of the irrep. 
$\{\vec{k}_{1,2}\}^\star$ denotes all momenta related to $\vec{k}_{1,2}$ by a rotation in the $O_h$ group. 
The $[\vec{k}_1,\vec{k}_2]$ label on the $\pi\pi$ operator indicates that it was constructed from single-pion operators with momenta in $\{\vec{k}_{1}\}^\star$ and $\{\vec{k}_{2}\}^\star$.
We refer to Ref.~\cite{Dudek:2012gj} for further details and explicit values of $\mathcal{C}$.
The various combinations of $\vec{k}_1$ and $\vec{k}_2$ used in our $\pi\pi$ operator constructions are presented in Table~\ref{tab:k1k2}.


\begin{table}[htbp]
\centering
\caption{The two-pion operators for each $\vec{P}$. Example momenta $\vec{k}_1$ and $\vec{k}_2$ are shown; all momenta in $\{\vec{k}_1\}^\star$ and $\{\vec{k}_2\}^\star$ are summed over in Eq.~\eqref{eq:two-meson}. Swapping around $\vec{k}_1$ and $\vec{k}_2$ gives the same operators up to an overall phase.}\label{tab:k1k2}
\[
\begin{array}{|c|c|c|c|}
	\hline \vec{P} &  \vec{k}_{1} & \vec{k}_{2} & \Lambda^{(P)} \\
	\hline[0,0,0] & \begin{array}{c}
		\begin{array}{c}
			{[0,0,1]} \\
			{[0,1,1]} \\
			{[1,1,1]}
		\end{array}
	\end{array} & \begin{array}{c}
	\begin{array}{c}
		{[0,0,-1]} \\
		{[0,-1,-1]} \\
		{[-1,-1,-1]}
	\end{array}
	\end{array} & \begin{array}{c}
	\begin{array}{c}
		T_{1}^{-} \\
		T_{1}^{-} \\
		T_{1}^{-}
	\end{array}
	\end{array} \\
	\hline[0,0,1]  & \begin{array}{c}
		\begin{array}{c}
    		{[0,0,0]} \\
			{[0,-1,0]} \\
			{[-1,-1,0]} \\
			{[0,0,-1]}
		\end{array}
	\end{array} & \begin{array}{c}
	\begin{array}{c}
    		{[0,0,1]} \\
			{[0,1,1]} \\
			{[1,1,1]} \\
			{[0,0,2]}
	\end{array}
        \end{array} & \begin{array}{c}
	\begin{array}{c}
		A_{1} \\
		A_{1}, E_{2}, B_{1} \\
		A_{1}, E_{2}, B_{2} \\
		A_{1}
	\end{array}
	\end{array} \\
	\hline[0,1,1]  & \begin{array}{c}
	\begin{array}{c}
		{[0,0,0]} \\
		{[0,1,0]} \\
		{[-1,0,0]} \\
		{[1,1,0]} \\
		{[0,1,-1]}
	\end{array}
	\end{array} & \begin{array}{c}
	\begin{array}{c}
		{[0,1,1]} \\
		{[0,0,1]} \\
		{[1,1,1]} \\
		{[-1,0,1]} \\
	    {[0,0,2]}
	\end{array}
	\end{array} & \begin{array}{c}
	  \begin{array}{c}
		A_{1} \\
		B_{1} \\
		A_{1}, B_{2} \\
		B_{1}, B_{2} \\
		A_{1}, B_{1}
	\end{array}
	\end{array} \\
	\hline[1,1,1]  & \begin{array}{c}
		\begin{array}{c}
 		{[0,0,0]} \\
		{[1,0,0]} \\
		{[2,0,0]}
    	\end{array}
	\end{array} & \begin{array}{c}
		\begin{array}{c}
			{[1,1,1]} \\
			{[0,1,1]} \\
			{[-1,1,1]}
		\end{array}
		\end{array} & \begin{array}{c}
			\begin{array}{c}
				A_{1} \\
				A_{1} \\
				A_{1}
			\end{array}
		\end{array} \\
		\hline [0,0,2] & {[0,0,0]} & {[0,0,2]} & A_{1} \\
		\hline
	\end{array}\]
\end{table}

\subsection{Variational analysis}
\label{sec:gevp}

After computing the correlation function matrix 
\begin{equation}
C_{ij}(t)  = \langle\,\mathcal{O}_i(t+t_0) \, \mathcal{O}_j(t_0)\,\rangle ,
\end{equation}
where the source time $t_0$ is averaged on all time slices, the finite-volume spectrum can be extracted by solving the GEVP method, 
\begin{equation} 
\label{eq:eigen_eq}
	C(t)u_n(t, t_0) = \lambda_n(t, t_0) \, C(t_0) \, u_n(t, t_0),
\end{equation}
where $n$ labels the eigenstates, $\lambda_n$ and $u_n$ are the $n$'th eigenvalue and eigenvector. 
Instead of solving the GEVP at all time slices, which could cause variance in the eigenvectors at different time slices due to possible degeneracies and numerical uncertainties, we employ the fixed-eigenvector method.~\cite{Bulava:2009jb, Mahbub:2013ala, Kiratidis:2015vpa, Kiratidis:2016hda}. 
We note that $C_{ij}$ is a Hermitian matrix in the ensemble average and as such we work with the hermitian matrix $\frac{1}{2}(C_{ij}+C_{ij}^\dagger)$.
This method solves the GEVP at a fixed time $\tilde{t}$ and estimate the eigenvalues at other time slices from the eigenvectors at $\tilde{t}$ as follows
\begin{equation}
\lambda_n(t, t_0) = u^{n*}_i(\tilde{t}, t_0) \, C_{ij}(t) \, u^n_j(\tilde{t}, t_0) \, ,
\label{eq:lambdan}
\end{equation}
In the current calculation, we find the eigenvalues with different choices of $(t_0,\tilde{t}) \in [4,10]$ is rather stable. 
So we just take $(t_0,\tilde{t}) = (5,6)$.
The energies are then obtained by fitting $\lambda_n(t, t_0)$ to an exponential form
\begin{equation}
	\lambda_n(t, t_0) = A \, e^{-E_n(t-t_0)}.
\label{eq:onestatefit}
\end{equation}
where $E_n$ is the energy of the $n$-th state. We also tried the two-state fit $\lambda_n(t, t_0) = (1-A)e^{-E_n(t-t_0)} + Ae^{-E'(t-t_0)}$, where the second exponential captures the contributions of the higher energy states and allows us to fit the correlators at earlier time slices. However, we find the higher energy states make minimal contributions even at small Euclidean times. So we just use the one-state fit as in Eq.~(\ref{eq:onestatefit}). 
%
%
%
%

\subsection{Phase-shift, pole position and extrapolation formulas }
\label{sec:Phase_shifts-tool}

The scattering phase shift is determined using L\"uscher's formalism, which establishes a connection between the finite-volume spectra and the scattering phase shift in an infinite volume. 
The general formula for single-channel scattering can be written as: 
\begin{equation} \label{eq:onedim_delta}
	\det\left(M_{ln,l^\prime n^\prime}^{(\vec{P}, \Lambda, \mu)}(q^2)-\delta_{ll^\prime}\delta_{nn^\prime}\cot(\delta_l)\right)=0,
\end{equation}
where $M_{ln,l^\prime n^\prime}^{(\vec{P}, \Lambda, \mu)}(q^2)$ is a known function of which the explicit form is provided below, $\delta_l$ is the phase shift. 
The parameter $q$ is defined in term of the scattering momentum $k$ as $q = k L / 2\pi$, and $k$ is related to finite-volume center-of-momentum frame energy $E^*=\sqrt{m_1^2 + k^2} + \sqrt{m_2^2 + k^2}$, with $m_{1,2}$ denoting the masses of the two scattering particles. 
In principle,  $M_{ln,l^\prime n^\prime}^{(\vec{P}, \Lambda, \mu)}(q^2)$ is an infinite-dimensional matrix with rows and columns indexed by $\{ln\}$ and $\{l^\prime n^\prime\}$, where $l(l^\prime)$ is the angular momentum subduced to the irrep $\Lambda$ and $n(n^\prime)$ is the $n(n^\prime)$'th embedding of that $l(l^\prime)$ into this irrep. 
For the isospin-1 $\pi\pi$ scattering, it is expected that $\delta_{l>3} \ll \delta_3 \ll \delta_1$. 
In this work we will only consider the $p$-wave and ignore the higher partial wave contributions. 
In this case, Eq.~\eqref{eq:onedim_delta} simplifies to
\begin{equation}
	\delta_1 = \arccot	M_{11,11}^{\left(\vec{P}, \Lambda,\mu\right)},
\end{equation}
It is convenient to define the notation $w_{j, s}$ as in Ref~\cite{ExtendedTwistedMass:2019omo}
\begin{equation}
	w_{j, s}=\frac{\mathcal{Z}_{j s}\left(1, q^2\right)}{\pi^{3 / 2} \sqrt{2 j+1} \gamma q^{j+1}},
\end{equation}
where $\mathcal{Z}_{j s}$ is the zeta-function, and $\gamma = \sqrt{E^{*2} + \vec{P}^2}/E^*$. 
The explicit expressions for $M_{11,11}^{\left(\vec{P}, \Lambda,\mu\right)}$ are listed in Tab.~\ref{tab:M1111}.

\begin{table}[htbp]
	\centering
	\caption{The function $M_{11,11}^{\left(\vec{P}, \Lambda,\mu\right)}$  for all momenta $\vec{P}$ and irreps $\Lambda$ used in this work. For $\vec{P}=(0,0,2)$, the functions are the same as $\vec{P} = (0,0,1)$, so they are omitted in the table. }\label{tab:M1111}
	\[
	\begin{array}{clll}
		\hline \hline 
		\vec{P} & \Lambda,\mu & M_{11,11}^{\left(\vec{P}, \Lambda,\mu\right)} \\
		\hline 
		(0,0,0) & T_1^-,2 & w_{0,0} \\ 
		(0,0,1) & A_1,1  & w_{0,0}+2  w_{2,0} \\
		(0,0,1) & E_2,1  & w_{0,0}-w_{2,0} \\
		(0,0,1) & E_2,2  & w_{0,0}-w_{2,0} \\
		(0,1,1) & A_1,1  & w_{0,0}+\frac{1}{2}w_{2,0}+i\sqrt{6}w_{2,1}-\frac{\sqrt{6}}{2}w_{2,2} \\
		(0,1,1) & B_1,1  & w_{0,0}+\frac{1}{2}w_{2,0}-i\sqrt{6}w_{2,1}-\frac{\sqrt{6}}{2}w_{2,2} \\
		(0,1,1) & B_2,1  & w_{0,0}-w_{2,0}+\sqrt{6}w_{2,2} \\
		(1,1,1) & A_1,1  & w_{0,0} - i2\sqrt{6}w_{2,2} \\
		\hline
		\hline
	\end{array}\]
\end{table}

The resulting phase shift is related to the relativistic BW form for the elastic $p$-wave amplitude in the resonance region~\cite{ParticleDataGroup:2024cfk},
\begin{equation}
	a_{1}=\frac{-\sqrt{s} \Gamma(s)}{s-m_{\rho}^{2}+\mathrm{i} \sqrt{s} \Gamma(s)}=\mathrm{e}^{\mathrm{i} \delta(s)} \sin \delta(s),\label{eq:BW}
\end{equation}
where $s=E^{*2}$ is the Mandelstam variable and $m_\rho$ is the peak position of resonance on the real axis. 
The decay width $\Gamma(s)$ is expressed in terms of the coupling constant $g_{\rho\pi\pi}$, taking into account the $\pi\pi$ phase space~\cite{Brown:1968zza,Renard:1974cz}
\begin{equation}
	\Gamma(s)=\frac{g_{\rho \pi \pi}^{2}}{6 \pi s}\left(\frac{s}{4}-m_\pi^2\right)^{3/2},
\end{equation}
%
Typically, the $\rho$ width $\Gamma_\rho = \Gamma(m_\rho^2)$ is evaluated at a phase shift of 90 degrees.

In Ref.~\cite{Bruns:2004tj} the pion mass dependence of the $\rho$-meson mass was computed using effective field theory with infrared regularization.
Neglecting $\mathcal{O}(m_\pi^3)$ and the non-analytic term of order $m_\pi^4$, considering the lattice spacing, the dependence reads
\begin{equation}
	m_\rho = c_0 + c_1 m_\pi^2 + c_2 a^2, \label{eq:extrapolation}
\end{equation}
We also consider the extrapolation of the coupling constant in a similar way, 
\begin{equation}
	g_{\rho\pi\pi} = \tilde{c}_0 + \tilde{c}_1 m_\pi^2 + \tilde{c}_2 a^2, \label{eq:extrapolation2}
\end{equation}

It is notable that the BW form is an assumption of the lineshape of the resonance.
In principle, the BW mass, which is $m_\rho$ in Eq.~(\ref{eq:BW}), should contain the contribution of hadronic loops.
Typically, for the $\rho$ meson case, the loop contributions of $\pi\pi$, $\pi\omega$, $K\bar{K}$ and even higher coupled channels have been absorbed into the BW form, which does not include the non-analytic quark mass dependence required by chiral symmetry.
Thus, to check the consistency of our two extrapolation equations, 
Eqs.~(\ref{eq:extrapolation}) and (\ref{eq:extrapolation2}), we next consider the loop contributions explicitly.

In particular, we use HEFT to re-analyze the lattice energy levels, in order to check whether the stability of the extracted pole position.
The workflow of the method based on HEFT is well developed in Refs.~\cite{Wu:2014vma,Hall:2013qba,Li:2019gya,Li:2021mob}, and we briefly introduce here. 
Initially, we need to construct the effective Hamiltonian of the system and impose the system in a finite volume with periodic boundary conditions. 
The parameters of the Hamiltonian are then calibrated by fitting its eigenvalues to the lattice spectra. 
Since the Hamiltonian in the finite and infinite volume involves the same parameters, once they are determined the observables in the infinite volume, such as phase shifts, can be obtained by solving the scattering equation.

In Ref.~\cite{Yu:2023xxf}, HEFT was employed to analyze the lattice spectra of the $\pi\pi\,,I=l=1$ sector, as provided by several different collaborations. 
In this work, we apply the same model to analyze our data. 
Specifically, the model includes a bare $\rho$ single-particle state, denoted as $\ket{\,\rho_B}$, and two-pion states and omega-pion states, denoted as $\ket{\,\pi\pi}$ and $\ket{\,\omega\pi}$, respectively. 
We will adopt two schemes, namely, A and B, to analyze the lattice data. 
In scheme A the $\ket{\,\omega\pi}$ channel is excluded, while in scheme B it is included. 
In both schemes we consider only the $P$-wave interaction between $\ket{\,\rho_B}$ and $\ket{\,\pi\pi},\ket{\,\omega\pi}$. 
The potential, which is parameterized in the center of mass frame, is specified by the following expression:
\begin{align}
V_{\rho\pi\pi}(k) &= \frac{g_{\rho\pi\pi} k}{\sqrt{m_{\rho}^B} E_\pi(k)} \left(\frac{\Lambda_{\rho\pi\pi}^2}{k^2+\Lambda_{\rho\pi\pi}^2}\right)^2 , 
\\
V_{\rho\omega\pi}(k) &= \frac{ g_{\rho\omega\pi} k\sqrt{m_\rho^B}}{\sqrt{2E_{\pi}(k)\,E_{\omega}(k)}} \left(\frac{\Lambda_{\rho\omega\pi}^2-\mu_\pi^2}{k^2 + \Lambda_{\rho\omega\pi}^2}\right)^2
\end{align}
with $E_{\pi/\omega}(k)=\sqrt{k^2+m_{\pi/\omega}^2}$. 
There are five parameters in the model: the bare rho mass, $m_{\rho}^B$, the coupling constants, $g_{\rho\pi\pi},\,g_{\rho\omega\pi}$, and the regulators,  $\Lambda_{\rho\pi\pi},\,\Lambda_{\rho\omega\pi}$. $\mu_\pi=138.5$ MeV is the physical pion mass.
Based on the previous work~\cite{Yu:2023xxf} and standard HEFT practice, the couplings and regulators are held fixed, independent of $m_\pi$ and the lattice spacing, while $m_\rho^B$ varies with both.

The resonance pole position is defined as the pole of scattering matrix $T_{\pi\pi\to\pi\pi}(z)$ on the second Riemann sheet. 
In this model, $T^{l=1}_{\pi\pi\to\pi\pi}(z)$ is given by
\begin{align}
	T^{l=1}_{\pi\pi\to\pi\pi}(z) = \frac{\abs{V_{\rho\pi\pi}(\bar{k}(z))}^2}{z-m_\rho^B - \Sigma(z)} ,
	\label{eq: tmatpipi2pipi}
\end{align}
where $\bar{k}(z) = \sqrt{\frac{z^2}{4}-m_\pi^2}$ and $\Sigma(z)=\Sigma_{\pi\pi}(z) + \Sigma_{\omega\pi}(z)$ denotes the self-energy correction defined as
\begin{align}
\Sigma_{\pi\pi}(z) &= \int q^2 dq  \frac{\abs{V_{\rho\pi\pi}(q)}^2}{z-2\sqrt{q^2+m_\pi^2} + i\epsilon},  \\
\Sigma_{\omega\pi}(z) &= \int q^2 dq  \frac{\abs{V_{\rho\omega\pi}(q)}^2}{z-\sqrt{q^2+m_\pi^2}-\sqrt{q^2+m_\omega^2} + i\epsilon} \quad . 
\end{align}
Therefore, the pole mass is obtained by solving the equation
\begin{align}
	z - m_\rho^B - \Sigma(z) = 0.
	\label{eq:pole equation}
\end{align}
We should mention that the real part of the $\Sigma(z)$ includes the non-analytic dependence on the quark mass. On the real-$z$ axis, the scattering matrix is related to the $\pi\pi$ $p$-wave phase shift in Eq.~(\ref{eq:BW}) as
\begin{align}
    \mathrm{e}^{i\delta(s)}\sin\delta(s) = -\rho(s) \,T^{l=1}_{\pi\pi\to\pi\pi}(\sqrt{s}) ,\label{eq:phaseHEFT}
\end{align}
where $\rho(s)=\frac{\pi}{4}\left(s\left(\frac{s}{4}-m_\pi^2\right)\right)^{1/2}$ comes from the phase space factor of $\pi\pi$.

To perform a chiral extrapolation and achieve the continuum limit, we parameterize the $m_\rho^B$ as follows:
\begin{align}
	m_\rho^B(m_\pi,a) = c_0 + c_1 m_\pi^2 + c_2 a^2 ,
	\label{eq:extrapolation to baremass}
\end{align}
where $a$ is the lattice spacing. 
Differing from Eq.~(\ref{eq:extrapolation}), it is important to note that Eq.~(\ref{eq:extrapolation to baremass}) is applied to the bare mass.
This is more physically reasonable as the bare mass is quark-model like.
The contributions of hadronic loops are separated in this framework.

\section{Results and discussion}
\label{sec:result}

\subsection{Finite volume spectra}
\label{sec:spectra}

\begin{figure*}[htbp]
	\centering
	\includegraphics[width=2\columnwidth]{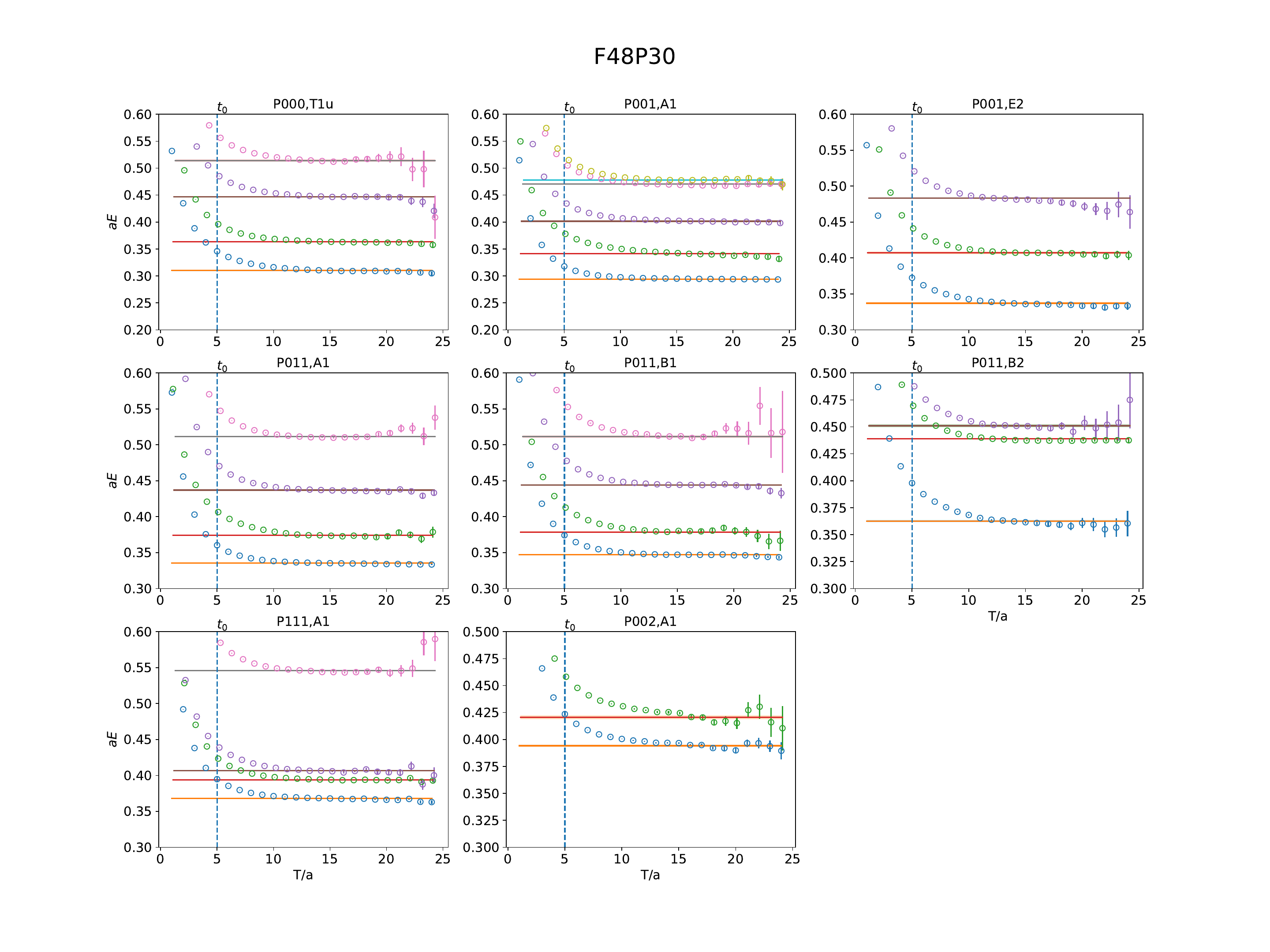}
\caption{Effective masses of the eigenvalues $\lambda_n(t)$ from the GEVP analysis for all $\vec{P}$ and $\Lambda$ on lattice F48P30. The horizontal lines indicate the fitted energies, $a_t E_n$.}
	\label{fig:lambda_F48P30}
\end{figure*}

For each group of ensembles, the GEVP approach provides the "principal correlators", $\lambda_n(t,t_0)$, of Eq.~(\ref{eq:lambdan}).
Following a comprehensive survey of various values for the parameters $t_0$ and $\tilde{t}$ in Eq.~(\ref{eq:lambdan}), we select $t_0=5$ and $\tilde{t}=6$.
As an example, in Fig.~\ref{fig:lambda_F48P30} we present the effective masses corresponding to the finite volume energies for all momenta and irreps computed on the ensemble F48P30. 
The presence of clear plateaus in all effective masses indicates that the energies can be extracted reliably with good precision.

\begin{figure*}[htbp]
	\centering
	\includegraphics[width=2\columnwidth]{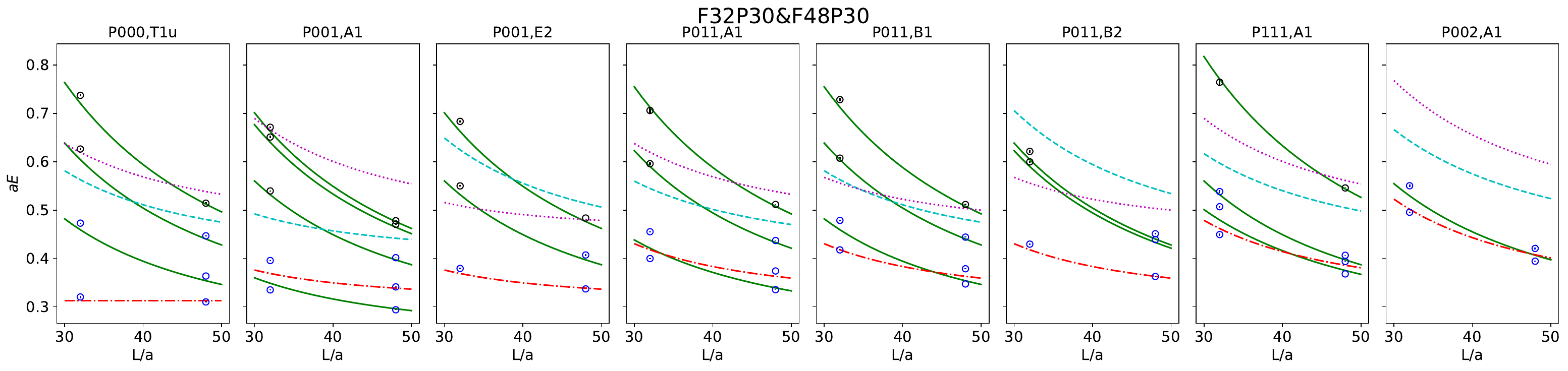}
	\caption{Energy spectra for all $\vec{P}$ and $\Lambda$ on lattices F32P30 and F48P30. The blue and black points show the extracted spectra. The green solid lines indicate the non-interacting $\pi\pi$ energies. The inelastic thresholds corresponding to the opening of the $K\bar{K}$ and $\pi\omega$ channels are indicated by the cyan dashed lines and purple dotted lines, respectively. The red dash-dotted line represents the mass calculated from the single particle operator. The black data points are not used in the scattering analysis because they are above the inelastic thresholds.}
	\label{fig:spec_F48P30}
\end{figure*}

The energy results for the ensembles F32P30 and F48P30, which share the same pion mass and lattice spacing, are shown in Fig.~\ref{fig:spec_F48P30}. 
The inelastic thresholds for the $K\bar{K}$ and $\pi\omega$ channels are also included in the figures. 
The multi-particle thresholds like $4\pi$, $\eta\pi\pi$ and $\pi K\bar{K}$ lie even higher and are not shown in the figure. 
Our scattering analysis will be restricted in the elastic region. 
Additional figures for other ensembles can be found in Appendix~\ref{sec:fig_spectra}, where the corresponding energy values are also provided.

From these spectra, we discover interesting phenomena that deepen our understanding of the finite-volume energy spectrum of the $\rho$.
Firstly, we find that regardless of whether it is in the rest frame or the moving frame, there is only one energy level between the two free $\pi\pi$ levels. 
This is because between the two free levels in the finite volume in the energy region under consideration, the scattering phase shift of $\pi\pi$ and the zeta function, which extends from negative infinity to positive infinity in the L\"uscher formula, are both monotonically increasing and have only one intersection point. 

Secondly, in the rest frame, there is one energy level far below the free energy of the lowest $\pi\pi$ state for all ensembles except C48P23 and C48P14. We estimate that the mass of the $\rho$ meson is actually below the energy of the lowest $p-$wave $\pi\pi$ state in these finite volumes. In other words, the $\rho$ meson cannot decay into the $\pi\pi$ final state. 

For the two ensembles C48P23 and C48P14, there is an energy between the free energies of the $\pi\pi$ states with momenta $|\vec{k_1}| = |\vec{k_2}| = \frac{2\pi}{L}$ and $|\vec{k_1}| = |\vec{k_2}| = \sqrt{2} \frac{2\pi}{L}$ -- see Figs.~\ref{fig:spec_C32P23_C48P23} and  \ref{fig:spec_C48P14}. 
This energy level aligns with the naive mass obtained from the single particle operator (in the following we will call it the single particle mass), which is indicated by the red dash-dotted line in the figures. 
We conjecture that the $\rho$ meson mass in these two ensembles is higher than the free energy of the lowest $p-$wave $\pi\pi$ state.
On the other hand, in the moving frame, as the minimum relative center-of-mass kinetic energy of $\pi\pi$ decreases, for example, for a system with a total momentum of $(0,0,1)$, the minimum momenta of $\pi\pi$ are $(0,0,0)$ and $(0,0,1)$ respectively, and the center-of-mass momentum is actually $(0,0,1/2)$, which is lower than the minimum center-of-mass momentum $(0,0,1)$ in the rest frame. 
Therefore, in these moving systems, the mass of the $\rho$ meson is more likely to be higher than the lowest free energy of $\pi\pi$. 
Under the lattice ensemble F48P30, as shown in Fig.~\ref{fig:spec_F48P30}, many energy levels in the moving systems are close to the single particle mass and lie between the two free energy levels of $\pi\pi$. 
This also reflects that in these moving systems, it is actually easier to obtain information about the decay of $\rho$ mesons.

Lastly, we find that all energy levels follow the same pattern, with most energy levels close to the single particle mass being far from the $\pi\pi$ free energies, and energy levels farther from the single particle mass being near the $\pi\pi$ free energies. 
If they are higher than the single particle mass, they are above the nearest free $\pi\pi$ energy, while if they are lower than the single particle mass, they are below the nearest free $\pi\pi$ energy.
These phenomena are consistent with second order perturbative mixing and the $\pi\pi$ phase shifts in this energy region. This will be discussed further in the next section.

\subsection{Phase shift and pole position}
\label{sec:Phase_shifts}

\begin{figure*}[htbp]
	\centering
	\includegraphics[width=0.9\columnwidth]{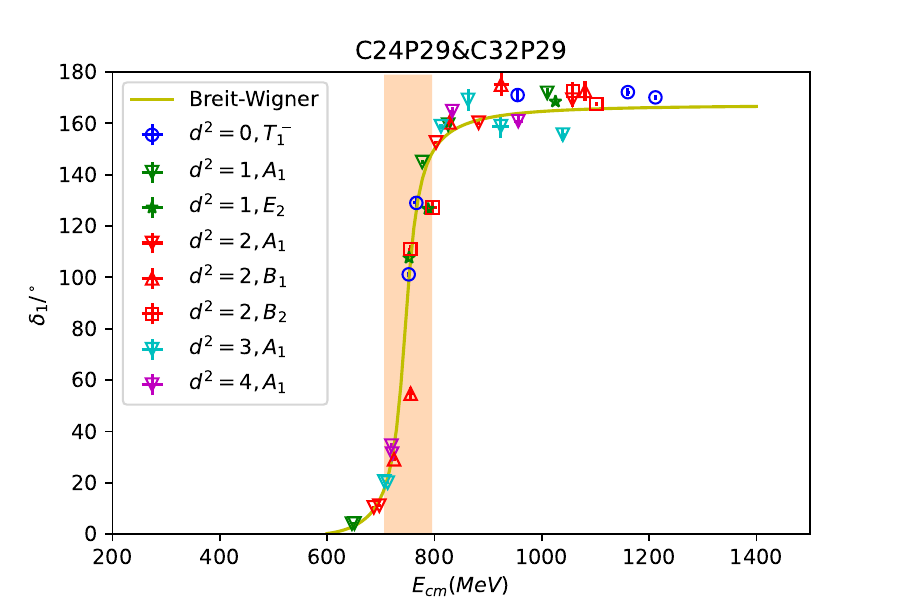}
	\includegraphics[width=0.9\columnwidth]{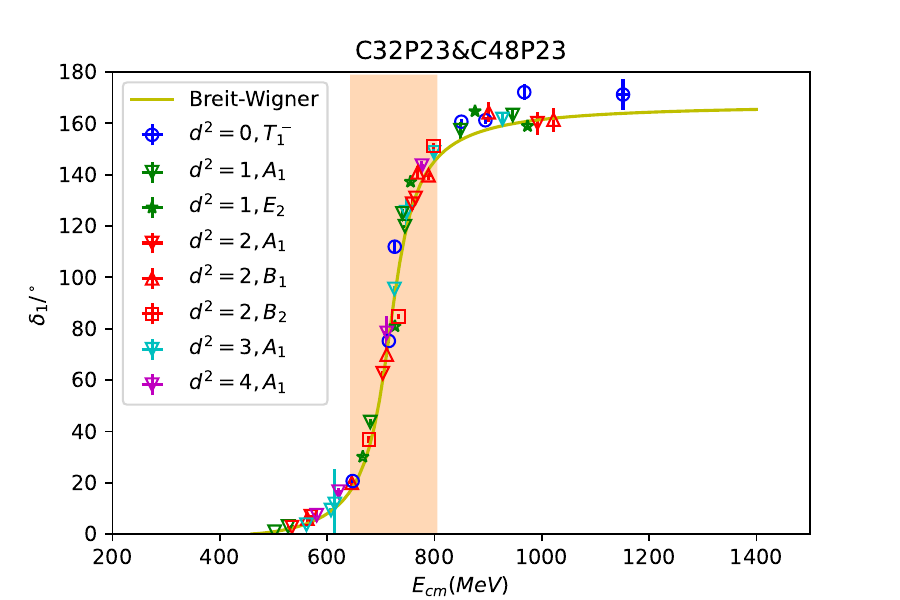} \\
	\includegraphics[width=0.9\columnwidth]{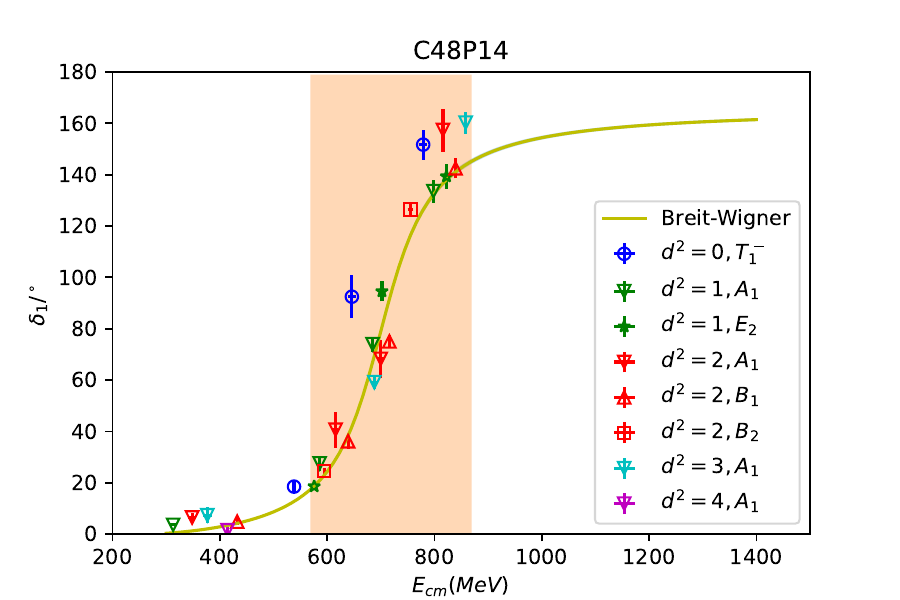}
	\includegraphics[width=0.9\columnwidth]{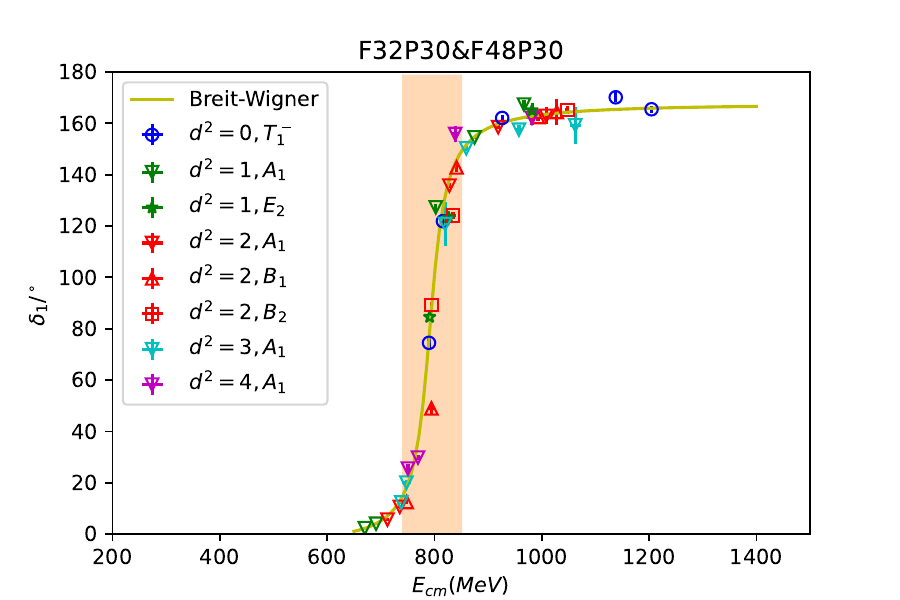} \\
	\includegraphics[width=0.9\columnwidth]{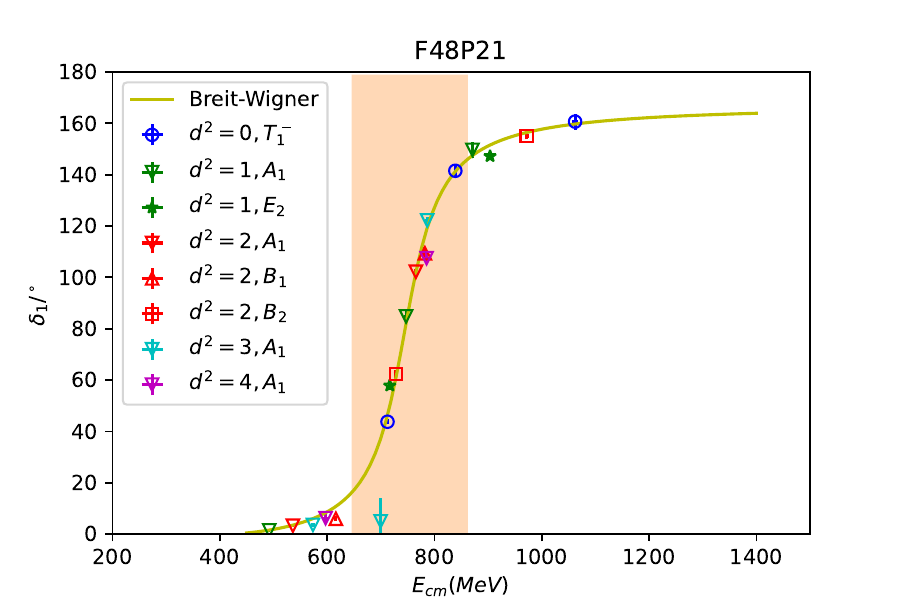}
	\includegraphics[width=0.9\columnwidth]{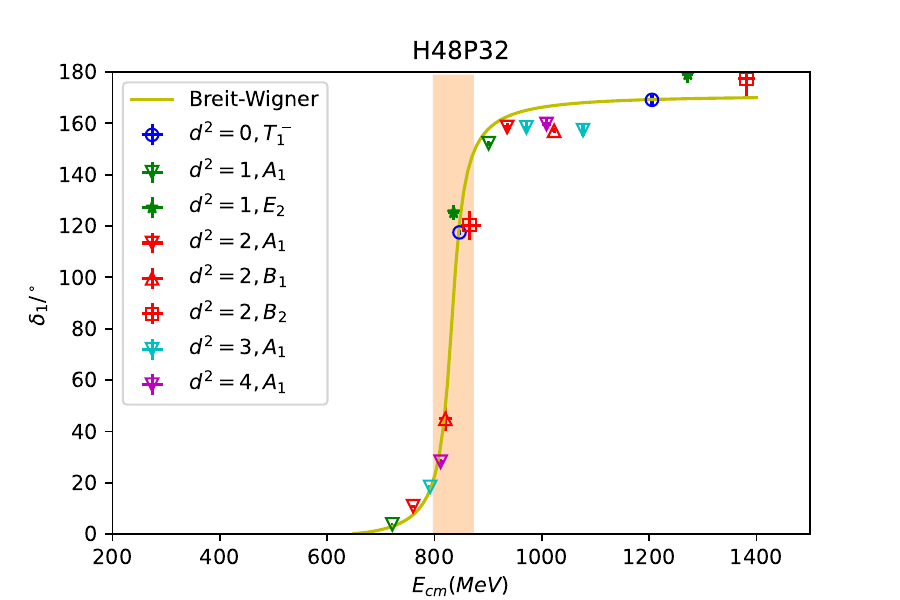}
	\caption{$P$-wave $\pi\pi$ elastic scattering phase shift $\delta_1$ as a function of the center-of-mass energy $(E_{cm})$. The curves represent the fit to the BW form. The vertical band highights the data used to constrain the HEFT analysis.}\label{fig:fit_delta}
\end{figure*}

In Fig.~\ref{fig:fit_delta} we present the determined phase shifts assuming that all $\delta_{l \geq 3}$ are negligible throughout the elastic region. 
We put the phase shifts of tuples (C24P29, C32P29), (C32P23, C48P23) and (F32P30, F48P30) together since the only difference between them is the volume. 
Two different volumes can offer more energy levels for fitting the scattering parameters, and also allow for the investigation of finite-volume corrections in the L\"ushcer's formula which scales as $\sim e^{-m_\pi L}$. 
For all three tuples mentioned, although the lineshapes of the phase shift from large lattice sizes are even more smooth, we did not observe significant differences between the two volumes, indicating that the finite-volume effect is under control in our results. 

For all the pion masses and lattice spacings we investigated, the phase shift jumps from 0 degrees to 180 degrees. 
This behavior implies the presence of a resonance associated with the $\rho$ meson in all scenarios. 
Furthermore, the energy corresponding to a phase shift around 90 degrees aligns with the BW mass of the $\rho$ meson.

We employ the BW form shown in 
Eq.~(\ref{eq:BW}) to fit the phase shifts obtained from the lattice spectra around the $\rho$ mass. 
The fitting results are given in Table~\ref{tab:mrho_g} and plotted in Fig.~\ref{fig:fit_delta}. 
The data points in the band for each case are fitted since the phase shift data far from the $\rho$ mass region suffer from the direct interaction of $\pi\pi\to \pi\pi$.
As evident from the figures, the fit describes the data well, particularly in the region where $\delta_1$ passes through $\pi/2$. 
The decomposition of $J=1$ into several irreps of the little groups on lattice is a consequence of symmetry breaking caused by the finite lattice spacing. 
Therefore, the difference between the irreps reflects discretization artifacts. 
We perform the analysis for each individual irrep and incorporate the difference between the irreps as a systematic uncertainty in our error estimation.


\begin{table}[]
\centering
\caption{Fit results of the BW form for $m_\rho$, $g_{\rho\pi\pi}$ and $Z_{\text{pole}}$.}\label{tab:mrho_g}
\resizebox{1\columnwidth}{!}{
\begin{tabular}{cccc}
\hline \hline
                  & $m_\rho$ (MeV)       & $g_{\rho\pi\pi}$ & $Z_{\text{pole}}$(MeV)\\
\hline
C24P29/C32P29       & 749.0(6.0)  & 5.859(79)  & $745.7(6.0)  - i\,19.7(1.5)$ \\
C32P23/C48P23       & 723.9(5.6)  & 5.827(14)  & $717.1(5.6)  - i\,37.2(2.8)$\\
C48P14              & 714.0(34.5) & 6.300(440) & $699.2(34.4) - i\,71.4(22.5)$\\
F32P30/F48P30       & 795.5(4.2)  & 5.775(58)  & $791.7(4.3)  - i\,22.6(1.0)$\\
F48P21              & 754.2(3.0)  & 5.954(38)  & $744.7(3.0)  - i\,49.5(1.7)$\\
H48P32              & 834.8(7.4)  & 4.916(45)  & $832.5(7.4)  - i\,18.1(0.9)$\\
\hline\hline
\end{tabular}
}
\end{table}

As discussed before, there is another method to extract the pole position of the $\rho$ meson by considering the hadronic loop contribution in HEFT.
In this approach, we employ two schemes, as discussed earlier.
In scheme A, $g_{\rho\pi\pi}$ and $\Lambda_{\rho\pi\pi}$ are fitted as $7.0\pm0.1$ and $0.9\pm0.1$ GeV, respectively, while various values of $m^B_\rho$ are extracted for extrapolation latter. 
In scheme B, $g_{\rho\pi\pi}$ and $\Lambda_{\rho\pi\pi}$ are fitted as $7.4\pm0.1$ and $1\pm0.1$ GeV, while $g_{\rho\omega\pi}$ and $\Lambda_{\rho\omega\pi}$ are fixed at $18\,\text{GeV}^{-1}$ and $1$ GeV. The latter  two values are constrained by the physical decay width of $\omega\to 3\pi$.
The mass of the $\omega$ for each ensemble is extracted by the quark bilinear operator for simplicity. It is consistent with the recent result determined from three pion scattering~\cite{Yan:2024gwp}.
Again the value of $m^B_\rho$ is fitted for each ensemble for extrapolation later.

Here, we choose the energy level data which generate the phase shifts in the band in Fig.~\ref{fig:fit_delta} to constrain our model parameters.
After fitting, we find that the energy levels in the finite volume are described very well, even for the energy levels not included in the fit.
The fits are shown in the figures of Appendix ~\ref{app:HEFT}. 

\begin{figure}[]
	\centering
	\includegraphics[width=\columnwidth]{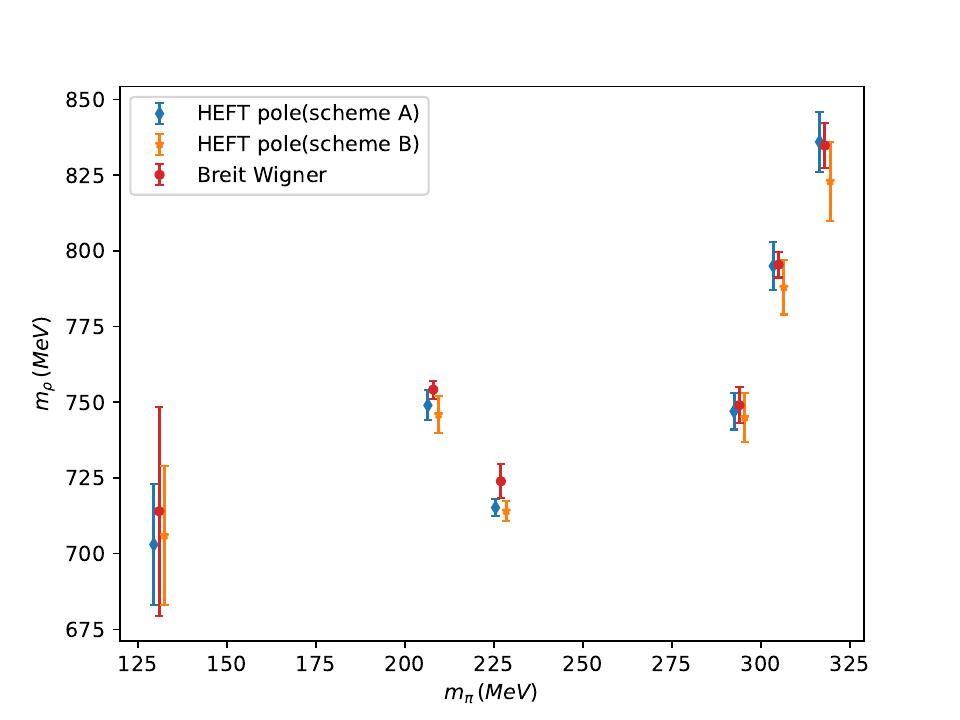}
	\caption{ The real part of the complex pole mass determined by HEFT in scheme A (blue solid diamond) and scheme B (yellow solid star). The red solid circles denote the BW mass present in Table~\ref{tab:mrho_g}. The HEFT pole positions are displaced horizontally for clarity.  }
	\label{fig:pole mass and bw mass}
\end{figure}

Then we can extract the pole positions of each ensemble from schemes A and B, to compare with those from the BW form.
In Fig.~\ref{fig:pole mass and bw mass}, we present the real part of the pole mass at each value of $m_\pi$ and compare it with the mass obtained using the BW parameterization. 
Within errors, the pole positions of the two methods exhibit excellent agreement.

\subsection{Extrapolations}

\begin{figure}[]
	\centering
	\includegraphics[width=\columnwidth]{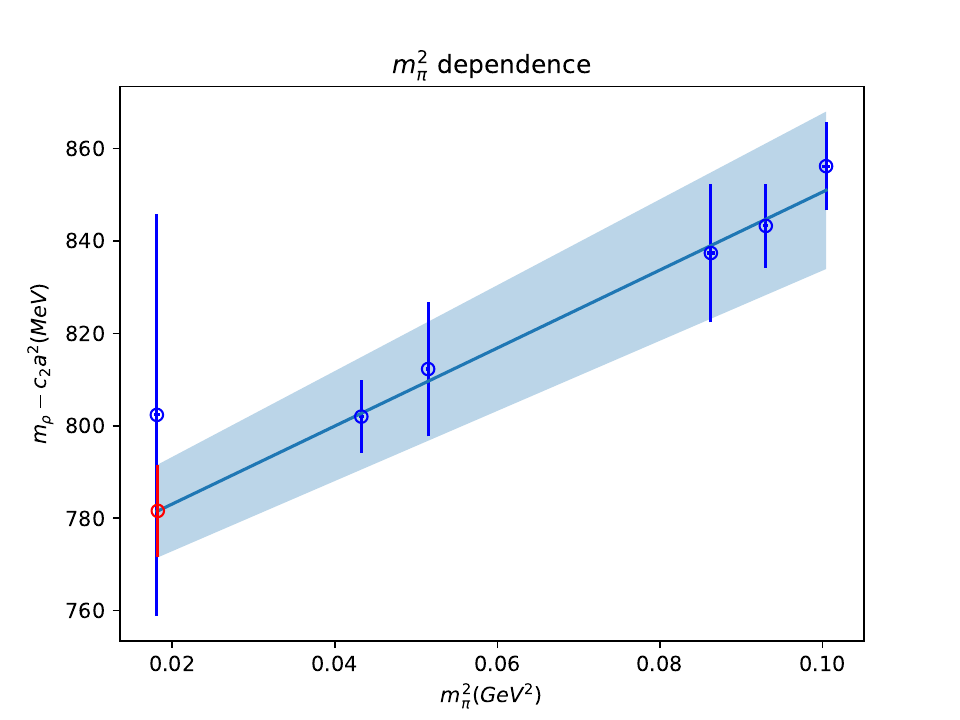}
	\caption{The result of extrapolation of $m_\rho$ using Eq.~(\ref{eq:extrapolation}).  
    The blue open circles denote $m_\rho$ in the BW form with elimination of lattice spacing artefacts parameterized by $c_2 a^2$. 
    The solid lines with gray error band denote the linear function $c_0 + c_1 m_\pi^2$.
    The red point is the result at the physical point.
    }\label{fig:same_a}
\end{figure}

We now have six solutions for the $\rho$ meson at different pion masses and lattice spacings.
Now we have two independent methods to do the extrapolation from the various pion mass and lattice spacing cases.
In the BW form, we extrapolate the parameters $m_{\rho}$ and $g_{\rho\pi\pi}$ to the physical pion mass and continuum limit using Eqs.~(\ref{eq:extrapolation}) and (\ref{eq:extrapolation2}), respectively. 
For $m_{\rho}$, the three parameters in the fitting formula are determined as follows: (the units are (MeV, GeV$^{-1}$, GeV$\cdot$ fm$^{-2}$))
\begin{align}
    (c_0,c_1,c_2)^{\text{BW}} &= (\,766.2 (8.5),\, 0.84(9),\, -7.97(81)\,)
\end{align}
with the $\chi^2/d.o.f=0.23$. 
For the coupling constant $g_{\rho\pi\pi}$, the parameters are found to be: (the units are (None, GeV$^{-2}$, fm$^{-2}$))
\begin{align}
(\tilde{c}_0,\tilde{c}_1,\tilde{c}_2)^{\text{BW}} &= (\,5.74 (67),\, -5.3 (6.4),\, 46 (48)\,),
\end{align}
with the $\chi^2/d.o.f=0.25$. Fig.~\ref{fig:same_a} shows $c_0+c_1m_\pi^2$.

Based on the above fitting results, we are able to obtain the mass and width of the $\rho$-meson at the physical pion mass $m_{\pi} = 135$~MeV and continuum limit as follows, 
\begin{equation}
	m_\rho = 781.6(10.0)\, \text{MeV}, \quad \Gamma_\rho(m_\rho) = 146.5(9.9)\, \text{MeV},
    \label{eq:BWpole}
\end{equation}
where width is calculated by the input of coupling constant $g_{\rho\pi\pi}=5.64(79)$.
These values are in good agreement with the experimental data~\cite{ParticleDataGroup:2024cfk}, although the uncertainties are still much larger than experimental data.

Using the two parameters $m_\rho$ and $g_{\rho\pi\pi}$ in Eq.~(\ref{eq:BW}), we can obtain the pole position of the $\pi\pi$ scattering T-matrix on the second Riemann sheet of the complex energy plane as follows, 
\begin{equation}
	Z_{\text{pole}}= 768.1\,(10.0)- i\,70.5\,(4.9)\, \text{MeV}.
\end{equation}

\begin{figure}[htbp]
	\centering
	\includegraphics[width=\columnwidth]{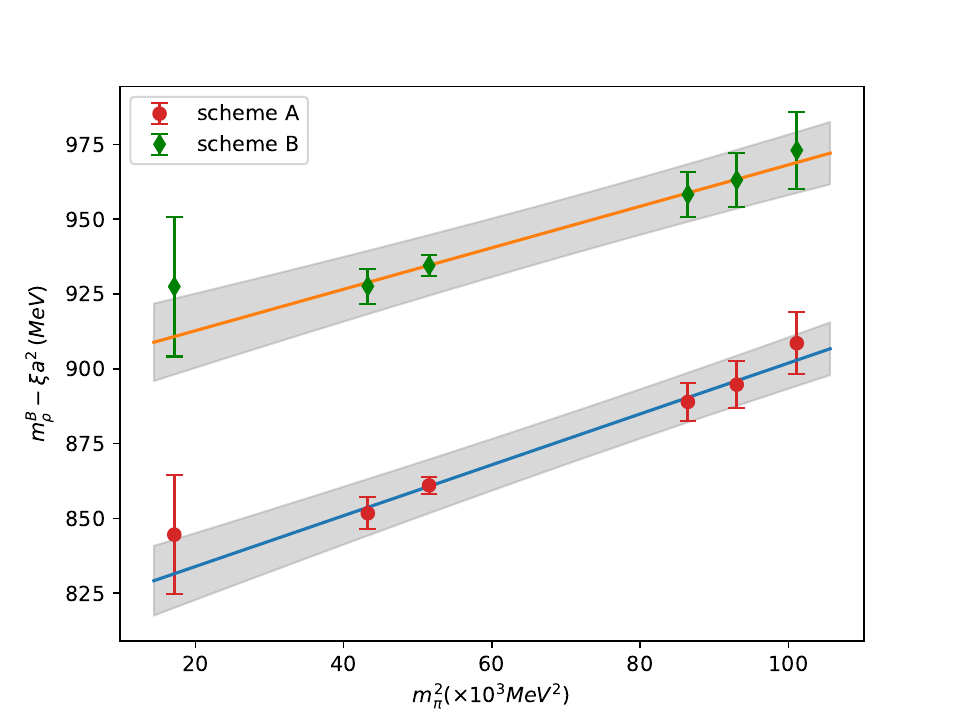}
	\caption{ 
    The same as Fig.~\ref{fig:same_a} but for the extrapolation of the bare $\rho$ mass, $m_\rho^B$, using 
    Eq.~(\ref{eq:extrapolation to baremass}).
     The red solid circles and green solid diamonds are for schemes A and B, respectively.
     The addition of the $\omega \pi$ channel in Scheme B requires a larger bare mass to fit the lattice QCD results.
    }
	\label{fig:mB extrapolation}
\end{figure}

In the framework of HEFT, we indeed extrapolate the bare mass of the $\rho$ meson, and the corresponding results are shown in Fig.~\ref{fig:mB extrapolation}. 
The central values and errors of $c_0,c_1,c_2$ for two fitting schemes are given by (the units are (MeV, GeV$^{-1}$, GeV$\cdot$ fm$^{-2}$))
\begin{align}
	(c_0,c_1,c_2)^{\text{A}} &= (\,817.0\,(13.0),\, 0.85\,(12),\, -7.61\,(89)\,), \\
    (c_0,c_1,c_2)^{\text{B}} &= (\,840.0\,(11.2),\, 0.72\,(11),\, -7.91\,(86)\,). 
\end{align}
Thus, we obtain $m_\rho^B(m_\pi^{\text{phy}},0) = 833\pm 13$ MeV for scheme A and $913\pm 13$ MeV for scheme B. 
The pole mass at the physical pion mass, $m_\pi^{\text{phy}}$, and the continuum limit are determined by solving Eq.~(\ref{eq:pole equation}) with the determined parameters. 
%
At last, combing with the error from the bare mass and other parameters, the final result is
\begin{align}
    m_\rho^{\text{pole,ext(A)}} = 787.0\,(15.0) - i\,59.0\,(6.0) \text{MeV},
    \\
    m_\rho^{\text{pole,ext(B)}} = 777.0\,(15.0) - i\,60.0\,(6.0) \text{MeV}. \label{eq:HEFTpole}
\end{align}
The extrapolated results are consistent with each other and both of them are also in agreement with the Particle Data Group value. 
It is also interesting to find that by including the $\omega\pi$ coupled channel the real part of the pole mass is reduced and becomes even closer to the experimental data.

At last, the pole masses extracted from the BW form and the HEFT method are consistent within a $1\sigma$ confidence interval.
Furthermore, to compare with the BW mass and width, we also extract the mass and width from the denominator of the T-matrix, as defined in.~(\ref{eq: tmatpipi2pipi}) at $\delta=90$ degree for the scheme B. 
The relationship between phase shift and T-matrix is shown in Eq.~(\ref{eq:phaseHEFT}) in this scenario. 
Then we obtain, 
\begin{equation}
	m_\rho = 782.0(13.5)\, \text{MeV}, \quad \Gamma_\rho(m_\rho) = 155.0(12.0)\, \text{MeV}.
    \label{eq:BWpole2}
\end{equation}
These values are in good agreement with the results obtained from the BW form, as shown in Eq.~(\ref{eq:BWpole}).
This consistent result ensures the reliability of our findings and indirectly reflects that when $m_\pi^2 < 0.1$ GeV$^2$, both the bare mass of the $\rho$ and the BW mass exhibit a linear relationship with the square of the $\pi$ mass. 
This is also consistent with the results in Ref.~\cite{Allton:2005fb}, as shown in Fig.~3 there.
Actually, that figure shows that at $m_\pi^2 \sim 0.175$ GeV$^2$, the linear relationship is significantly disrupted. 
%
%
%
%

Furthermore, we consider the uncertainties arising from the different mass of the strange quark in these configurations.
We add an additional term $c_4 (m^2_K-m^2_{\pi})$ in Eq.~(\ref{eq:extrapolation}) to re-fit the data, where $m_K$ is the mass of $K$ meson read from Tab.~\ref{tab:lattice}.
However, because of the small variation in the mass of the $K$ meson in these ensembles,  fitting four parameters with six data points becomes challenging.
Therefore, adding an additional variable to the fit does not allow for the detection of the dependence of the $\rho$ meson mass on the strange quark mass. 
This necessitates the need for more ensembles corresponding to different $\pi$ masses, $K$ masses, and lattice spacings in the future in order to truly investigate the dependence of the $\rho$ meson mass on all three light quark masses.

\section{Summary}
\label{sec:Summary}

\begin{figure*}[htbp]
	\centering
	\includegraphics[width=2.35\columnwidth]{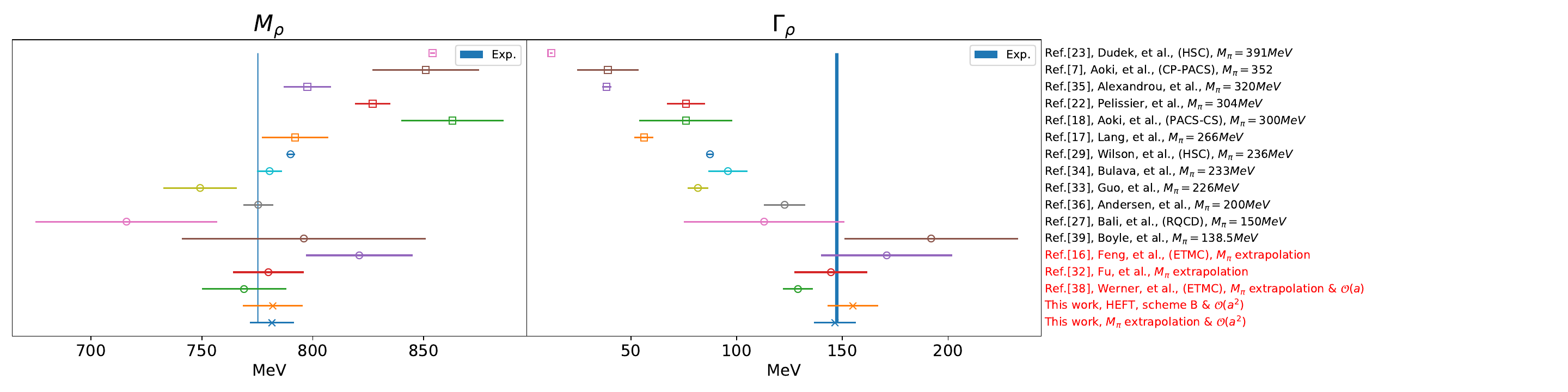}
	\caption{Comparison between different LQCD-based calculations of the $\rho$ mass and width. Among them only four made an extrapolation in $m_\pi$, which are marked red. In addition, only two made a lattice spacing extrapolation. In the present work, we offer two results based on the BW assumption and the HEFT framework, which both include a correction for finite lattice spacing up to $\mathcal{O}(a^2)$. To ensure consistency with other studies, we present the masses and widths of two methods from Eq.~(\ref{eq:BWpole}) and Eq.~(\ref{eq:BWpole2}), respectively. 
    }\label{fig:rhocompare}
\end{figure*}

Using nine $N_f = 2 + 1$ Wilson-Clover ensembles with three lattice spacings and various pion masses in the range of 135 to 320 MeV, we computed the mass and width of the $\rho$ resonance and successfully extrapolated the results to the physical pion mass and continuum limit. 
These results agree well with the experimental measurements. 
While lattice QCD studies of the ground state hadron spectra have already achieved great success, calculations of resonance properties at the physical point that can be directly compared with experiment are still scarce.
This study has demonstrated that, even for resonances that are unstable under strong interactions, accurate physical results for the spectral parameters can be obtained through a systematic, state-of-the-art lattice QCD investigation. 

A large number of operators, including the quark bilinear operators and $\pi\pi$ operators, are constructed at both rest frame and moving frames with total momenta $\vec{P}$ = $(0,0,1), (0,1,1), (1,1,1)$ and $(0,0,2)$, in units of $\frac{2\pi}{L}$.
The finite-volume energies are then extracted from the correlation function matrices of these operators. 

Assuming that partial waves with $l\geq 3$ are negligible, we determine the phase shifts using L\"uscher's finite volume method. 
At all of the lattice spacings and pion masses that we investigated, the energy dependence of the phase shifts shows an increasing tendency from 0 to 180 degrees, indicating the presence of a resonance. 
The mass and width of the resonance are then obtained by fitting the phase shifts to the relativistic BW form and these are further extrapolated to the physical pion mass and continuum limit. 
The extrapolated values of the BW mass and width, $(m_\rho,\,\Gamma_\rho) = (781.6\pm10.0,\, 146.5\pm 9.9)$ MeV, are in good agreement with the corresponding experimental values.

To further examine our result, we employ an alternative method of analysis based on HEFT. This method manifestly includes the quark mass dependence of the hadronic loop contributions of the $\pi\pi$ and $\omega\pi$ channels and extrapolates the bare mass of the $\rho$ meson.
The pole position of the $\rho$ meson at the physical pion mass and continuum limit determined through this method is consistent with the results using the BW form.
Thus, it does appear that our results are consistent and thus the properties of the $\rho$ meson have been successfully extracted.

We compare our results with the previous measurements of $\rho$ meson spectral parameters from lattice simulations in Fig.~\ref{fig:rhocompare}.
Based on these comparisons, it is seen that, thanks to the large number of ensembles at various lattice spacings and pion masses and a relatively complete operator set for each ensemble, this work presents the most precise lattice determination to date for the mass and width of the $\rho$ meson at the physical pion mass and in the continuum limit.

\section*{Acknowledgments}
We thank helpful discussions with Ying Chen, Feng-kun Guo, Keh-Fei Liu, Yibo Yang, Ross Young, James Zanotti, and Bing-Song Zou. 
We thank the CLQCD collaborations for providing us their gauge configurations with dynamical fermions, which are generated on HPC Cluster of ITP-CAS, the Southern Nuclear Science Computing Center(SNSC), the Siyuan-1 cluster supported by the Center for High Performance Computing at Shanghai Jiao Tong University and the Dongjiang Yuan Intelligent Computing Center.
This work is supported by the National Natural Science Foundation of China under Grant Nos. 12175239, 12221005, 12293060, 12293061, 12293063, 11935017 and 12175279, 
and by the Chinese Academy of Sciences under Grant No. YSBR-101.
It was also supported by the Australian Research Council through Grant Nos.\ DP210103706 (DBL) and DP230101791 (AWT).

L.~Liu also acknowledges support from the Strategic Priority Research Program of the Chinese Academy of Sciences with Grant No. XDB34030301 and Guangdong Major Project of Basic and Applied Basic Research No. 2020B0301030008.

\begin{appendices}

\section{Fitting results for HEFT}
\label{app:HEFT}

\begin{figure*}[htbp]
	\centering
	\includegraphics[width=2\columnwidth]{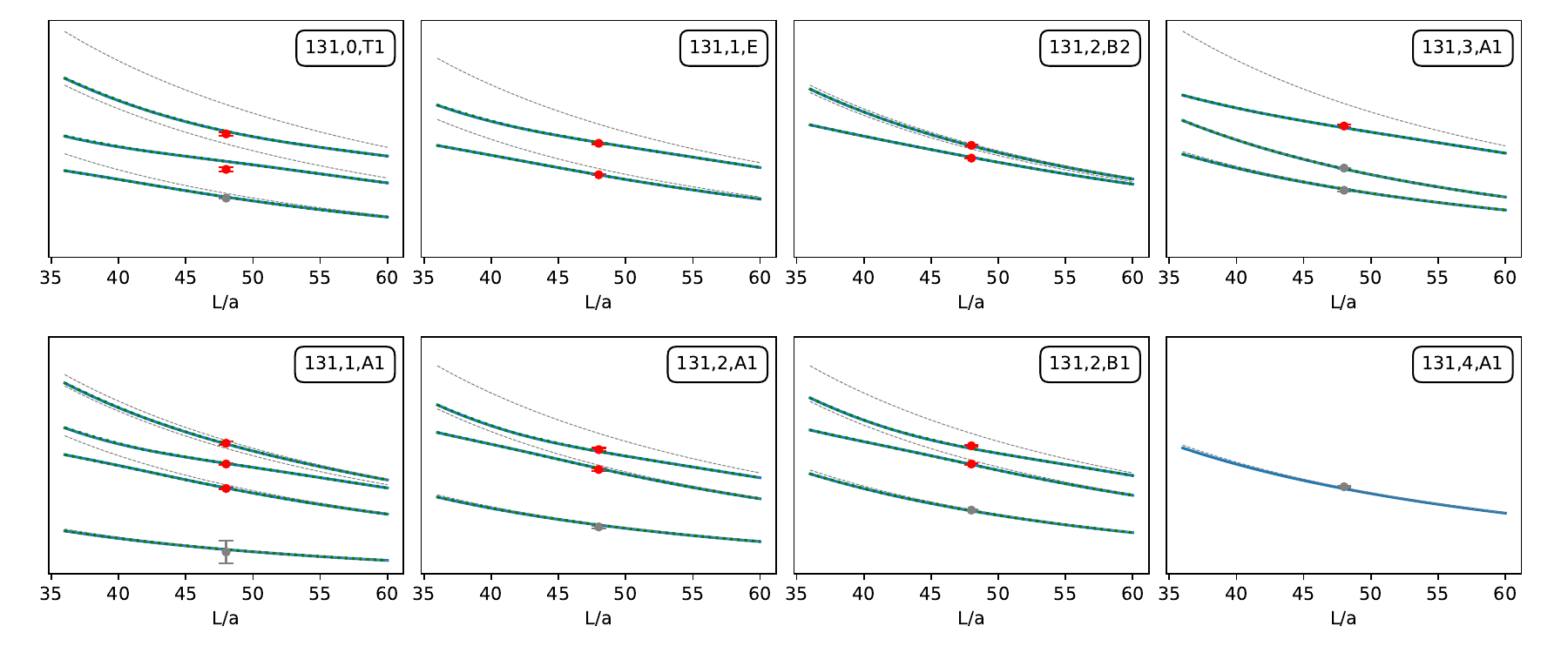}
	\includegraphics[width=2\columnwidth]{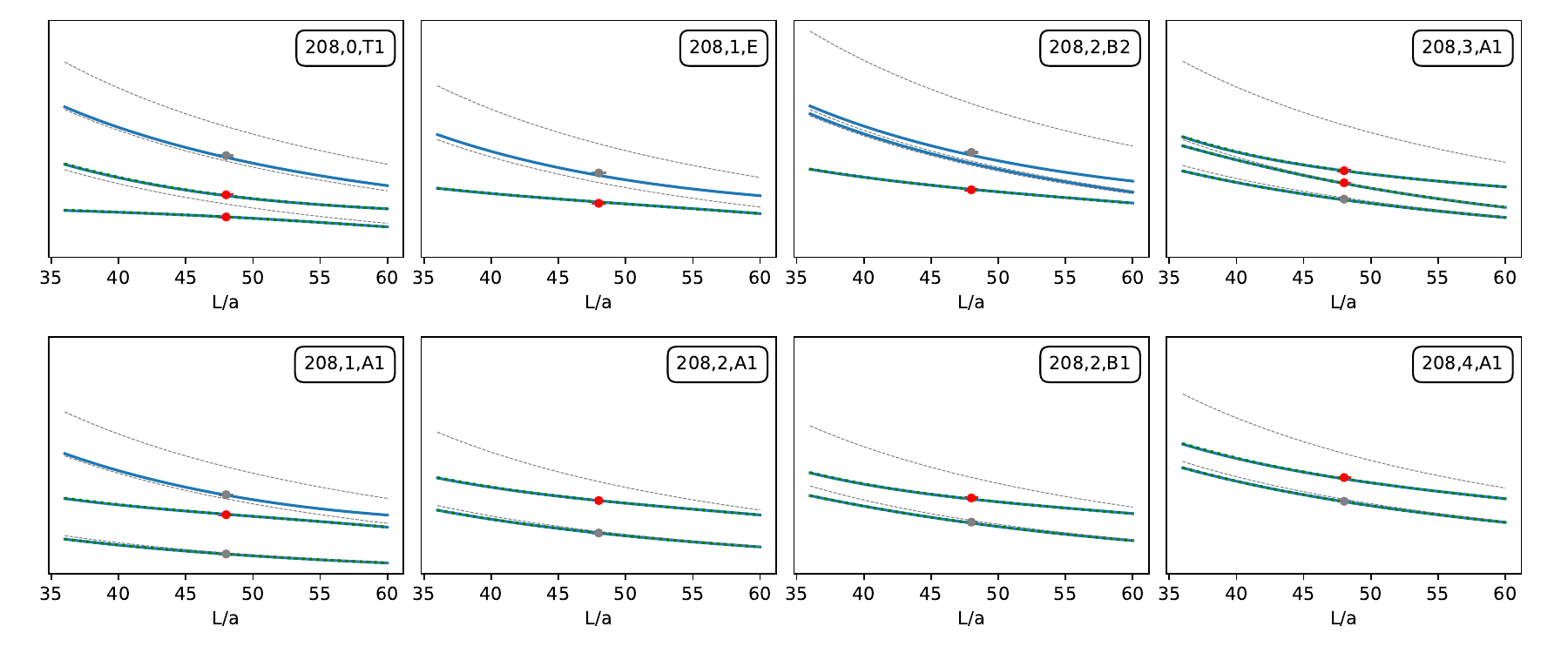}
	\includegraphics[width=2\columnwidth]{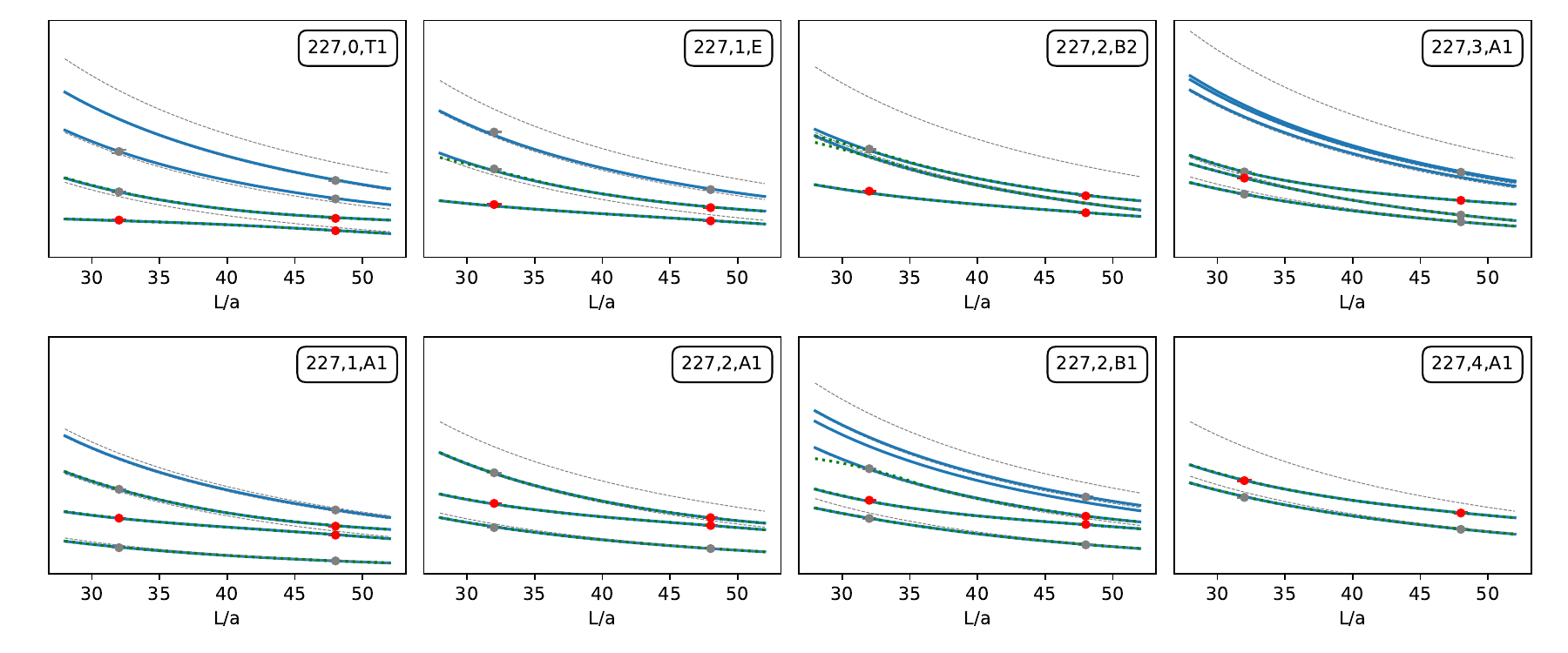}
	\caption{ The fit to lattice spectra using HEFT. The data points denote the lattice spectra in the present work. The red points are the energy levels lying in the band in Fig.~\ref{fig:fit_delta}, which are used for fitting. The gray points are not used for analysis. The bold, often overlapping blue solid and green dotted lines denote the eigenvalues of the finite volume Hamiltonian in schemes A and B, respectively. The gray dotted lines denote the free, non-interacting energy levels of $\pi\pi$. The annotation in the upper right corner denotes $m_\pi$, $(\frac{\vec{P}L}{2\pi})^2$, with $\vec{P}$ the total momentum of the system, and the irreducible representation of octahedral group. }
	\label{fig:HEFTfit1}
\end{figure*}

\begin{figure*}[htbp]
	\centering
	\includegraphics[width=2\columnwidth]{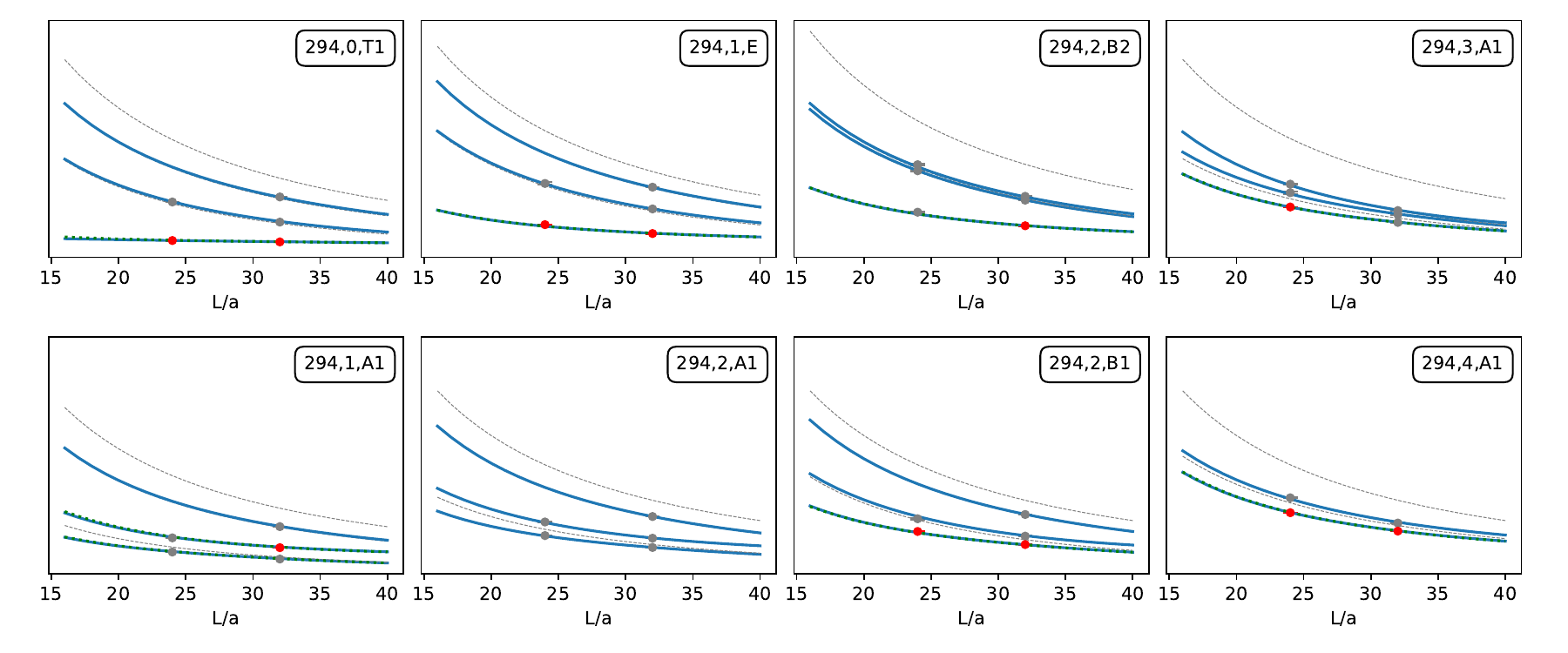}
	\includegraphics[width=2\columnwidth]{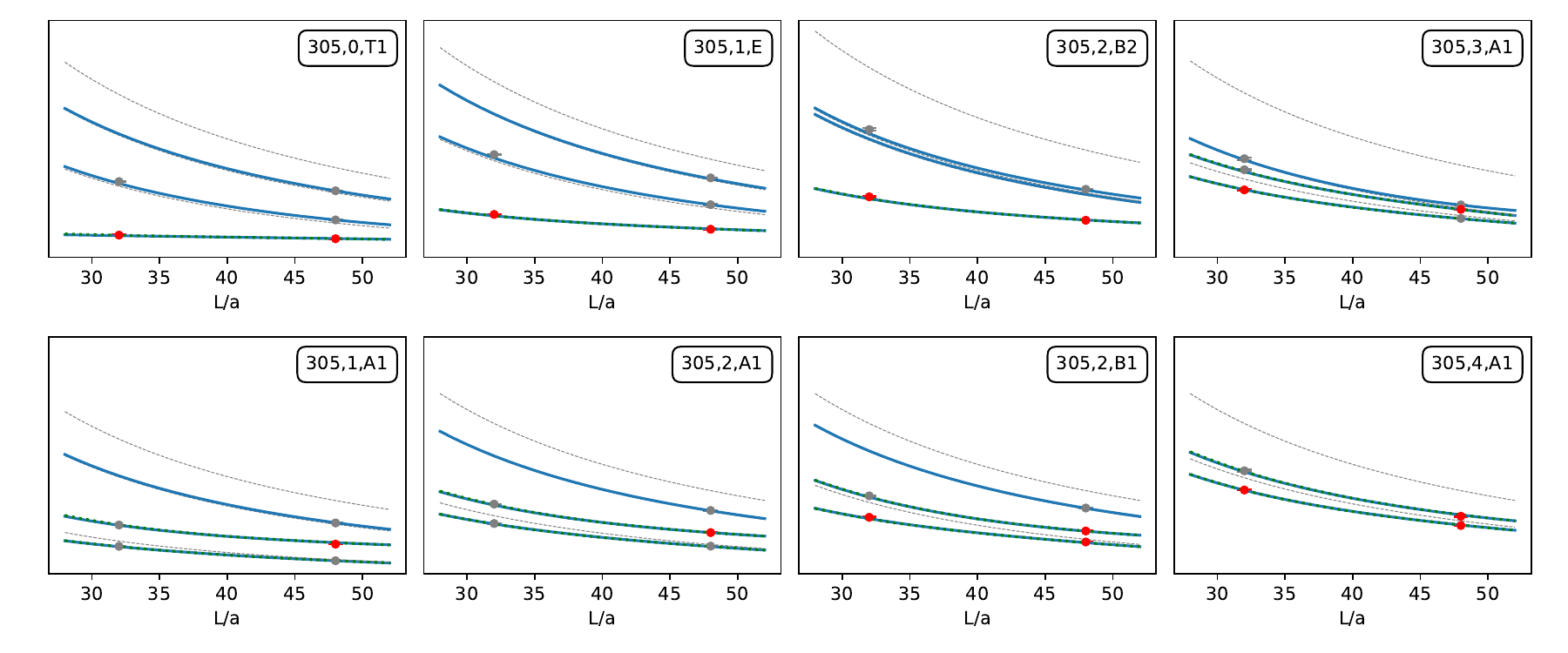}
	\includegraphics[width=2\columnwidth]{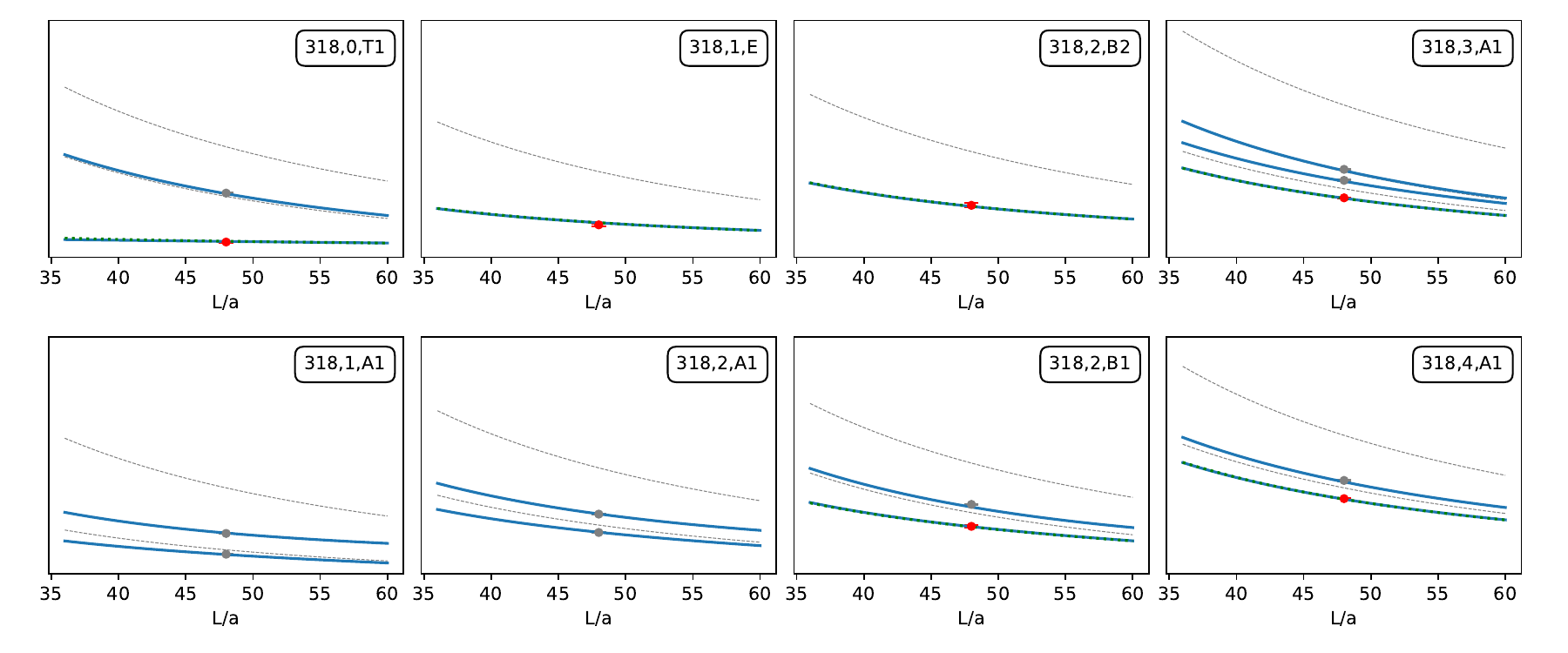}
	\caption{Same as for Fig.\ref{fig:HEFTfit1}}
	\label{fig:HEFTfit2}
\end{figure*}

The fitting results based on HEFT are shown in Figs.~\ref{fig:HEFTfit1} and \ref{fig:HEFTfit2} for schemes A and B, respectively.

\section{Lattice QCD Dataset}
\label{sec:fig_spectra}
\begin{figure*}[htbp]
	\centering
	\includegraphics[width=2\columnwidth]{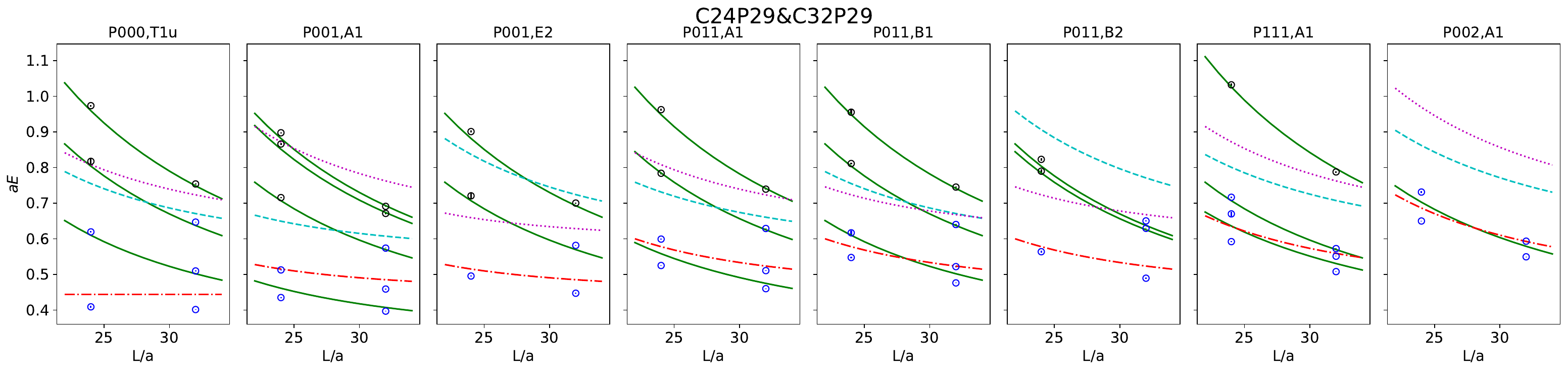}
	\caption{As for Fig.~\ref{fig:spec_F48P30} for C24P29 and C32P29.}
	\label{fig:spec_C24P29_C32P29}
\end{figure*}

\begin{figure*}[htbp]
	\centering
	\includegraphics[width=2\columnwidth]{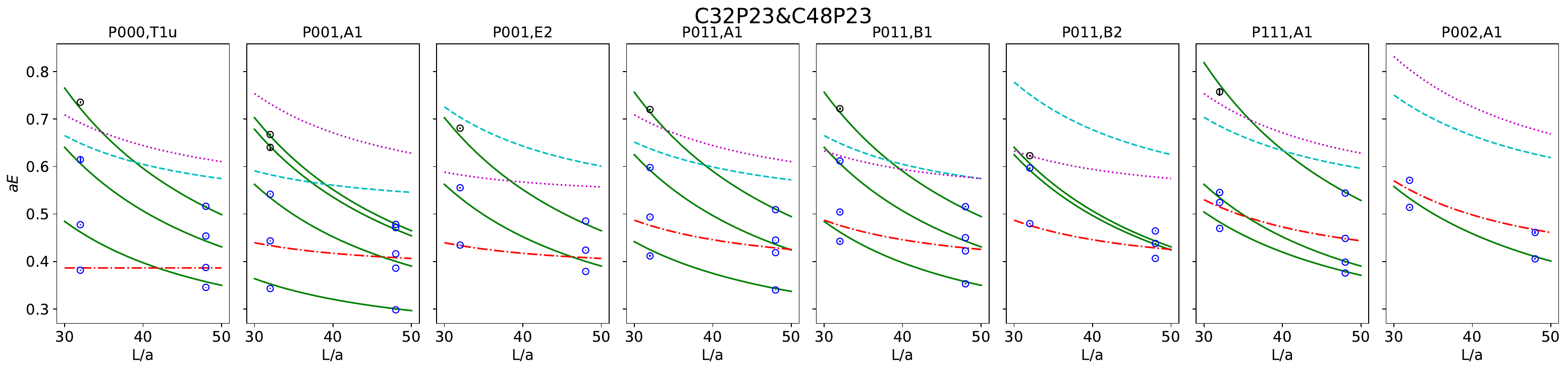}
	\caption{As for Fig.~\ref{fig:spec_F48P30} for C32P23 and C48P23.}
	\label{fig:spec_C32P23_C48P23}
\end{figure*}

\begin{figure*}[htbp]
	\centering
	\includegraphics[width=2\columnwidth]{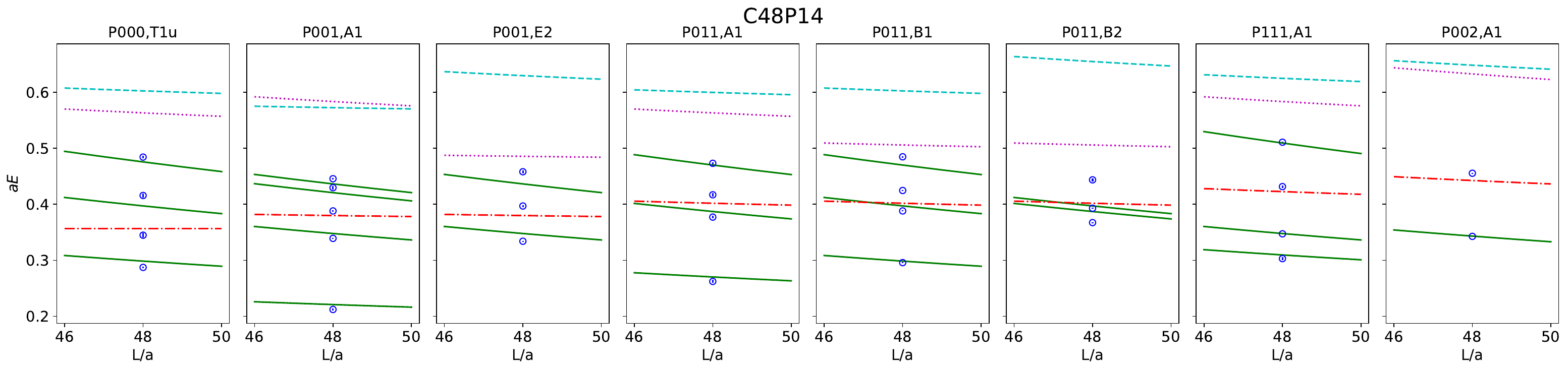}
	\caption{As for Fig.~\ref{fig:spec_F48P30} for C48P14.}
	\label{fig:spec_C48P14}
\end{figure*}

In Figs.~\ref{fig:spec_C24P29_C32P29}-\ref{fig:spec_H48P32} we show the finite volume spectra for the ensembles C24P29\&C32P29, C32P23\&C48P23, C49P14, F48P21 and H48P32. The energy values for all $(\vec{P}, \Lambda)$ and all ensembles used in this work are given in Tables~\ref{tab:data000T1}-\ref{tab:data002A1}.

\begin{table*}[htbp]
	\centering
	\caption{Energy levels of $\vec{P}=[000],\Lambda=T_1^-$}\label{tab:data000T1}
	\resizebox{2\columnwidth}{!}{
		\begin{tabular}{|l|ccccccccc|}
			\hline  
			& C24P29 &  C32P29 & C32P23 & C48P23 & C48P14 & F32P30  & F48P30  & F48P21 & H48P32 \\   
			\hline    
			$aE_0$ & 0.4089(32) & 0.4013(10) & 0.38155(99) & 0.34559(44) & 0.2898(12) & 0.3203(26) & 0.3101(13) & 0.27971(67) & 0.22257(87)  \\
			$aE_1$ & 0.6193(19) & 0.5096(15) & 0.4776(11) & 0.3874(11) & 0.36003(94) & 0.472911 & 0.36360(48) & 0.3293(10) & 0.3169(10) \\
			$aE_2$ & 0.8172(72) & 0.64674(91) & 0.6145(57) & 0.45371(70) & 0.4152(37) & 0.6261(22) & 0.4467(12) & 0.4170(20) & 0.4087(30) \\
			$aE_3$ & 0.9735(21) & 0.7538(13) & 0.7354(23) & 0.51625(87) & 0.48958(92) & 0.7370(19) & 0.5141(19) & 0.4965(14) & 0.4812(30) \\
			\hline  
		\end{tabular}
	}
\end{table*}	

\begin{table*}[htbp]
	\centering
	\caption{Energy levels of $\vec{P}=[001],\Lambda=A_1$}
	\resizebox{2\columnwidth}{!}{
		\begin{tabular}{|l|ccccccccc|}
			\hline  
			& C24P29 &  C32P29 & C32P23 & C48P23 & C48P14 & F32P30  & F48P30  & F48P21 & H48P32 \\   
			\hline    
			$aE_0$ & 0.4350(17) & 0.39685(86) & 0.3431(10) & 0.29826(54) & 0.2146(19) & 0.33493(64) & 0.29388(43) & 0.23336(58) & 0.23045(38)  \\
			$aE_1$ & 0.5125(20) & 0.45886(98) & 0.44343(77) & 0.38597(46) & 0.34060(77) & 0.39558(84) & 0.34111(86) & 0.3212(12) & 0.27062(88) \\
			$aE_2$ & 0.7155(16) & 0.57372(81) & 0.5417(10) & 0.41621(57) & 0.39052(62) & 0.5395(13) & 0.40149(56) & 0.3660(10) & 0.3630(18) \\
			$aE_3$ & 0.8660(27) & 0.6711(30) & 0.6403(57) & 0.47134(43) & 0.4325(21) & 0.6509(26) & 0.47046(74) & 0.4248(43) & 0.4336(18) \\
			$aE_4$ & 0.8975(16) & 0.6909(16) & 0.6675(16) & 0.47811(91) & 0.44639(96) & 0.6711(18) & 0.4779(10) & 0.4476(20) & 0.4487(14) \\
			\hline  
		\end{tabular}
	}
\end{table*}	

\begin{table*}[htbp]
	\centering
	\caption{Energy levels of $\vec{P}=[001],\Lambda=E_2$}
	\resizebox{2\columnwidth}{!}{
		\begin{tabular}{|l|ccccccccc|}
			\hline  
			& C24P29 &  C32P29 & C32P23 & C48P23 & C48P14 & F32P30  & F48P30  & F48P21 & H48P32 \\   
			\hline    
			$aE_0$ & 0.4954(16) & 0.4471(17) & 0.4345(11) & 0.37897(39) & 0.33435(76) & 0.3792(16) & 0.33699(74) & 0.3103(13) & 0.2557(23)  \\
			$aE_1$ & 0.7205(83) & 0.58143(80) & 0.5554(14) & 0.42390(48) & 0.40075(68) & 0.5499(14) & 0.4071(18) & 0.37811(80) & 0.3591(25) \\
			$aE_2$ & 0.9008(28) & 0.7002(13) & 0.6808(18) & 0.48525(55) & 0.4577(11) & 0.6832(19) & 0.48348(64) & 0.45796(83) & 0.4523(27) \\
			\hline  
		\end{tabular}
	}
\end{table*}	

\begin{table*}[htbp]
	\centering
	\caption{Energy levels of $\vec{P}=[011],\Lambda=A_1$}
	\resizebox{2\columnwidth}{!}{
		\begin{tabular}{|l|ccccccccc|}
			\hline  
			& C24P29 &  C32P29 & C32P23 & C48P23 & C48P14 & F32P30  & F48P30  & F48P21 & H48P32 \\   
			\hline    
			$aE_0$ & 0.52490(98) & 0.45991(85) & 0.4117(14) & 0.34003(54) & 0.26821(98) & 0.39975(87) & 0.33539(40) & 0.28030(66) & 0.27240(63)  \\
			$aE_1$ & 0.5990(14) & 0.51055(68) & 0.49366(76) & 0.41867(42) & 0.38129(79) & 0.4552(11) & 0.37401(74) & 0.35289(69) & 0.3078(14) \\
			$aE_2$ & 0.7838(31) & 0.62880(76) & 0.5978(13) & 0.44513(50) & 0.4196(23) & 0.5958(25) & 0.43697(61) & 0.40416(70) & 0.3976(26) \\
			$aE_3$ & 0.9623(32) & 0.7392(16) & 0.7201(14) & 0.50912(54) & 0.4739(12) & 0.7058(52) & 0.51152(65) & 0.4825(21) & 0.4869(13) \\
			\hline  
		\end{tabular}
	}
\end{table*}	

\begin{table*}[htbp]
	\centering
	\caption{Energy levels of $\vec{P}=[011],\Lambda=B_1$}
	\resizebox{2\columnwidth}{!}{
		\begin{tabular}{|l|ccccccccc|}
			\hline  
			& C24P29 &  C32P29 & C32P23 & C48P23 & C48P14 & F32P30  & F48P30  & F48P21 & H48P32 \\   
			\hline    
			$aE_0$ & 0.5474(25) & 0.47610(73) & 0.4426(21) & 0.35296(88) & 0.29742(75) & 0.4176(17) & 0.34722(37) & 0.30464(59) & 0.2842(23)  \\
			$aE_1$ & 0.6167(59) & 0.5218(18) & 0.5043(18) & 0.42221(39) & 0.38978(60) & 0.4786(16) & 0.3786(11) & 0.3586(17) & 0.32649(76) \\
			$aE_2$ & 0.8111(23) & 0.6400(12) & 0.6119(14) & 0.45015(53) & 0.42609(80) & 0.6073(28) & 0.4440(10) & 0.41244(85) & 0.4108(17) \\
			$aE_3$ & 0.9552(71) & 0.7450(18) & 0.7219(31) & 0.5154(14) & 0.48712(93) & 0.7282(35) & 0.5114(14) & 0.4908(11) & 0.4952(15) \\
			\hline  
		\end{tabular}
	}
\end{table*}	

\begin{table*}[htbp]
	\centering
	\caption{Energy levels of $\vec{P}=[011],\Lambda=B_2$}
	\resizebox{2\columnwidth}{!}{
		\begin{tabular}{|l|ccccccccc|}
			\hline  
			& C24P29 &  C32P29 & C32P23 & C48P23 & C48P14 & F32P30  & F48P30  & F48P21 & H48P32 \\   
			\hline    
			$aE_0$ & 0.5636(33) & 0.4893(18) & 0.4798(14) & 0.4065(10) & 0.36931(80) & 0.4295(17) & 0.36261(93) & 0.3405(14) & 0.2933(43)  \\
			$aE_1$ & 0.7897(46) & 0.62922(96) & 0.5968(17) & 0.43803(69) & 0.39404(80) & 0.5996(15) & 0.43896(53) & 0.40265(79) & 0.4042(12) \\
			$aE_2$ & 0.8228(37) & 0.6501(11) & 0.6229(19) & 0.46430(88) & 0.4454(16) & 0.6212(37) & 0.4510(14) & 0.4241(12) & 0.4076(40) \\
			\hline  
		\end{tabular}
	}
\end{table*}	

\begin{table*}[htbp]
	\centering
	\caption{Energy levels of $\vec{P}=[111],\Lambda=A_1$}
	\resizebox{2\columnwidth}{!}{
		\begin{tabular}{|l|ccccccccc|}
			\hline  
			& C24P29 &  C32P29 & C32P23 & C48P23 & C48P14 & F32P30  & F48P30  & F48P21 & H48P32 \\   
			\hline    
			$aE_0$ & 0.59192(87) & 0.50779(70) & 0.4697(13) & 0.37576(50) & 0.3087(14) & 0.4493(23) & 0.36795(46) & 0.31954(65) & 0.30780(64)  \\
			$aE_1$ & 0.6698(63) & 0.55097(77) & 0.5243(10) & 0.39860(80) & 0.34824(70) & 0.5069(16) & 0.39369(42) & 0.35617(75) & 0.3415(12) \\
			$aE_2$ & 0.7164(17) & 0.5724(12) & 0.5454(14) & 0.44863(79) & 0.4317(24) & 0.5381(31) & 0.40642(55) & 0.3831(11) & 0.36269(81) \\
			$aE_3$ & 1.0320(37) & 0.7875(23) & 0.7573(78) & 0.54409(50) & 0.51285(77) & 0.7641(84) & 0.54567(65) & 0.52156(84) & 0.5296(12) \\
			\hline  
		\end{tabular}
	}
\end{table*}	

\begin{table*}[htbp]
	\centering
	\caption{Energy levels of $\vec{P}=[002],\Lambda=A_1$}\label{tab:data002A1}
	\resizebox{2\columnwidth}{!}{
		\begin{tabular}{|l|ccccccccc|}
			\hline  
			& C24P29 &  C32P29 & C32P23 & C48P23 & C48P14 & F32P30  & F48P30  & F48P21 & H48P32 \\   
			\hline    
			$aE_0$ & 0.64977(98) & 0.54933(75) & 0.5140(16) & 0.4056(13) & 0.3425(11) & 0.4954(15) & 0.39418(85) & 0.3514(10) & 0.33770(90)  \\
			$aE_1$ & 0.7311(33) & 0.5932(14) & 0.5710(17) & 0.4611(30) & 0.4554(11) & 0.5502(30) & 0.4206(17) & 0.4044(13) & 0.3727(16) \\
			\hline  
		\end{tabular}
	}
\end{table*}

\begin{figure*}[htbp]
	\centering
	\includegraphics[width=2\columnwidth]{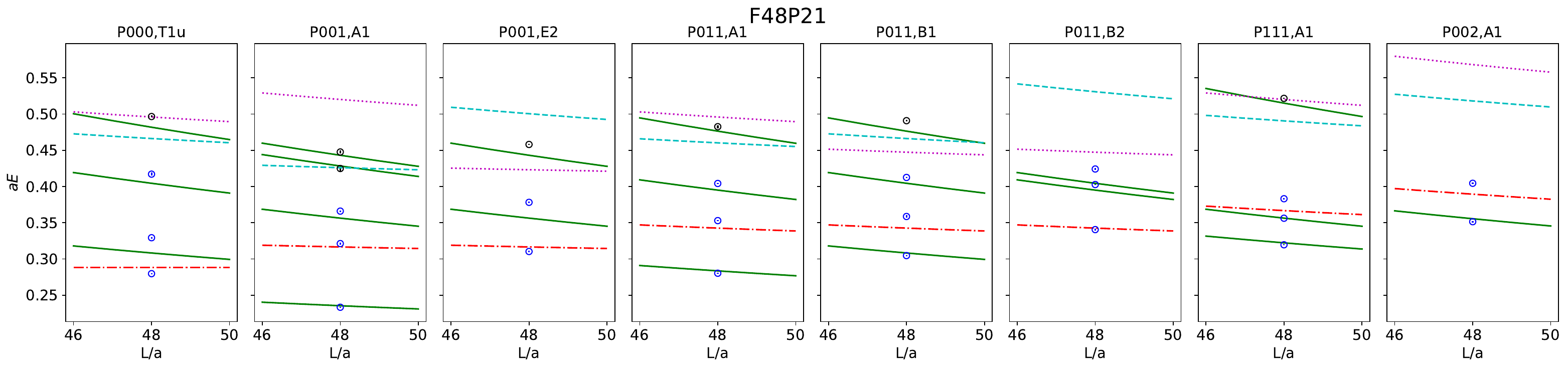}
	\caption{As for Fig.~\ref{fig:spec_F48P30} for F48P21.}
\end{figure*}

\begin{figure*}[htbp]
	\centering
	\includegraphics[width=2\columnwidth]{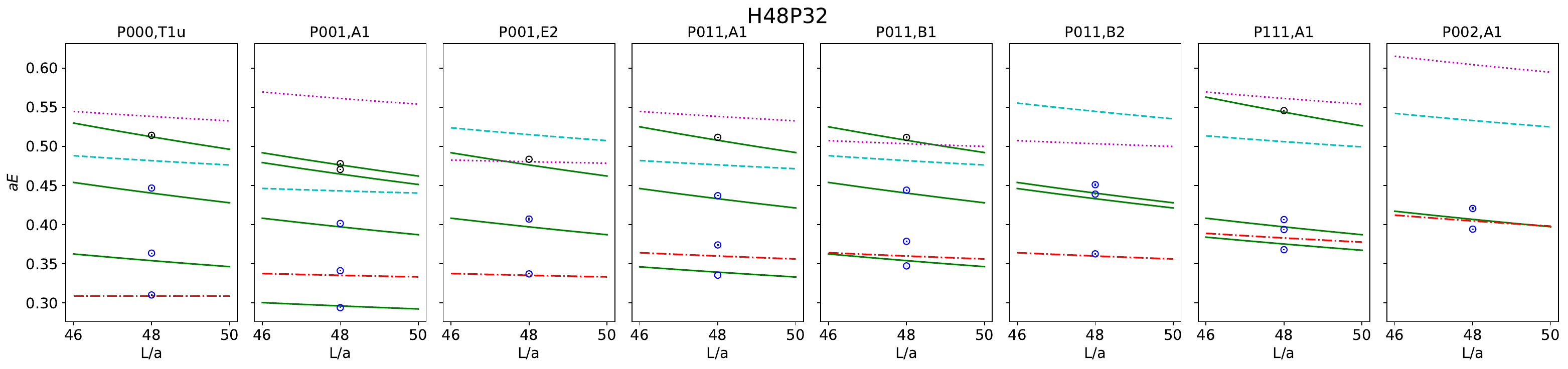}
	\caption{As for Fig.~\ref{fig:spec_F48P30} for H48P32.}
	\label{fig:spec_H48P32}
\end{figure*}

\end{appendices}


\bibliographystyle{elsarticle-num}
\bibliography{ref}

%


\end{document}